\documentclass{aa}  
\usepackage{txfonts}
\usepackage{natbib,twoopt}
\usepackage[breaklinks=true]{hyperref} 
\bibpunct{(}{)}{;}{a}{}{,}             
\makeatletter
  \newcommandtwoopt{\citeads}[3][][]{\href{http://adsabs.harvard.edu/abs/#3}%
    {\def\hyper@linkstart##1##2{}%
     \let\hyper@linkend\@empty\citealp[#1][#2]{#3}}}
  \newcommandtwoopt{\citepads}[3][][]{\href{http://adsabs.harvard.edu/abs/#3}%
    {\def\hyper@linkstart##1##2{}%
     \let\hyper@linkend\@empty\citep[#1][#2]{#3}}}
  \newcommandtwoopt{\citetads}[3][][]{\href{http://adsabs.harvard.edu/abs/#3}%
    {\def\hyper@linkstart##1##2{}%
     \let\hyper@linkend\@empty\citet[#1][#2]{#3}}}
  \newcommandtwoopt{\citeyearads}[3][][]%
    {\href{http://adsabs.harvard.edu/abs/#3}
    {\def\hyper@linkstart##1##2{}%
     \let\hyper@linkend\@empty\citeyear[#1][#2]{#3}}}
\makeatother

\usepackage{graphicx}   

\newcommand{\ie}{i.e.\xspace}
\newcommand{\eg}{e.g.\xspace}

\def\starlight{\textsc{starlight}\xspace}

\newcommand{\msun}{\ifmmode \text{M}_{\odot} \else M$_{\odot}$\fi\xspace}

\newcommand{\Mstar}{\ifmmode M_{\star} \else $M_{\star}$\fi\xspace}
\newcommand{\MBH}{\ifmmode M_\text{BH} \else $M_\mathrm{BH}$\fi\xspace}
\newcommand{\mdot}{\ifmmode \dot{m} \else \dot{m}\fi\xspace}
\newcommand{\Lrad}{\ifmmode L_\mathrm{1.4} \else $L_\mathrm{1.4}$\fi\xspace}
\newcommand{\Frad}{\ifmmode F_\mathrm{1.4} \else $F_\mathrm{1.4}$\fi\xspace}
\newcommand{\LHa}{\ifmmode L_{\Ha} \else $L_{\Ha}$\fi\xspace}
\newcommand{\Loiii}{\ifmmode L_{\oiii} \else $L_{\oiii}$\fi\xspace}
\newcommand{\Lbol}{\ifmmode L_\mathrm{bol} \else $L_\mathrm{bol}$\fi\xspace}
\newcommand{\Lbolmod}{\ifmmode L_\mathrm{bol}^\mathrm{mod} \else $L_\mathrm{bol}^\mathrm{mod}$\fi\xspace}
\newcommand{\Ledd}{\ifmmode L_\mathrm{Edd} \else $L_\mathrm{Edd}$\fi\xspace}
\newcommand{\Lmech}{\ifmmode L_\mathrm{mech} \else $L_\mathrm{mech}$\fi\xspace} 
\newcommand{\Avneb}{\ifmmode A_{V}^\mathrm{neb} \else $A_{V}^\mathrm{neb}$\fi\xspace} 
\newcommand{\Avstar}{\ifmmode A_{V}^\star \else $A_{V}^\star$\fi\xspace} 
\newcommand{\sigstar}{\ifmmode \sigma_{\star} \else $\sigma_{\star}$\fi\xspace} 

\newcommand{\hii}{\ifmmode \mathrm{H}\,\textsc{ii} \else H~{\sc ii}\fi\xspace}
\newcommand{\heii}{\ifmmode \mathrm{He}\,\textsc{ii} \else He~{\sc ii}\fi\xspace}
\newcommand{\Ha}{\ifmmode \mathrm{H}\alpha \else H$\alpha$\fi\xspace}
\newcommand{\Hb}{\ifmmode \mathrm{H}\beta \else H$\beta$\fi\xspace}
\newcommand{\Heii}{\ifmmode \mathrm{He}\textsc{ii} \else He~{\sc ii}\fi\xspace}
\newcommand{\oiii}{\ifmmode [\mathrm{O}\,\textsc{iii}] \else [O~{\sc iii}]\fi\xspace}
\newcommand{\oii}{\ifmmode [\mathrm{O}\,\textsc{ii}] \else [O~{\sc ii}]\fi\xspace}
\newcommand{\oi}{\ifmmode [\mathrm{O}\,\textsc{i}] \else [O~{\sc i}]\fi\xspace}
\newcommand{\nii}{\ifmmode [\mathrm{N}\,\textsc{ii}] \else [N~{\sc ii}]\fi\xspace}
\newcommand{\sii}{\ifmmode [\mathrm{S}\,\textsc{ii}] \else [S~{\sc ii}]\fi\xspace}
\newcommand{\Oiii}{[O~{\sc iii}]$\lambda$5007\xspace}
\newcommand{\Nii}{[N~{\sc ii}]$\lambda$6584\xspace}

\newcommand{\Opp}{O$^{++}$} 

\newcommand{\woiii}{\ifmmode W[\mathrm{O}\,\textsc{iii}] \else $W[\mathrm{O}\,\textsc{iii}]$\fi\xspace}
\newcommand{\wha}{\ifmmode W(\mathrm{H}\alpha) \else $W(\mathrm{H}\alpha)$\fi\xspace}

\titlerunning{Optically active and optically inactive radio galaxies}
\authorrunning{G. Stasi\'nska et al.}

\begin{document}

\title{Optically active and optically inactive radio galaxies\\ as sub-populations of the main galaxy sample of the SDSS}

\author{G. Stasi\'nska
        \inst{\ref{obspm}}
        \and
        N. Vale Asari
        \inst{\ref{ufsc}}
        \and
          A. W\'ojtowicz
        \inst{\ref{brno}}
                \and
        D. Kozie\l-Wierzbowska
        \inst{\ref{krakow}}
      }

\institute{LUTH, Observatoire de Paris, CNRS, Universit\'e Paris Diderot; Place Jules Janssen, F-92190 Meudon, France\label{obspm}\\
           \email{grazyna.stasinska@obspm.fr}
           \and
           Departamento de F\'{\i}sica--CFM, Universidade Federal de Santa Catarina, C.P.\ 5064, 88035-972, Florian\'opolis, SC, Brazil\label{ufsc}\\
           \email{natalia@astro.ufsc.br}
           \and
           Department of Theoretical Physics and Astrophysics, Faculty of Science, Masaryk University, Kotláršká2, Brno, 61137, Czech Republic\label{brno}\\
           \email{awojtowicz@oa.uj.edu.pl}
                      \and
           Astronomical Observatory, Jagiellonian University, ul. Orla 171, PL-30244 Krak\'ow, Poland\label{krakow}\\
           }

   \date{Received ???; accepted ???}
           

\abstract 
{}
{
We use the ROGUE I and II catalogues of radio sources associated with optical galaxies to revisit the characterization of radio active galactic nuclei (AGNs) in terms of radio luminosities and properties derived from the analyses of the optical spectra of their associated galaxies. 
}
{
We propose a physically based classification of radio galaxies into  `optically inactive'  and `optically active'  (OPARGs and OPIRGs). In our sample, there are 14082 OPIRGs and 2721 OPARGs.
After correcting for the Malmquist bias, we compared the global properties of our two classes of radio galaxies and put them in the context 
of the global population of galaxies. To compare the Eddington ratios of OPARGs with those of Seyferts, we devised a method to obtain the bolometric luminosities of these objects, taking into account the contribution of young stars to the observed line emission.  We provide formulae to derive bolometric luminosities from the \oiii luminosity. 
}
{
We find that  the distributions of radio luminosities of OPARGs and OPIRGs are  undistinguishable. On average, the black hole masses and stellar masses in OPIRGs are larger than in OPARGs.
OPARGs show signs of some recent star formation. 
Plotting the OPARGs in the BPT diagram and comparing their distribution with that of the remaining galaxies, we find that there is a sub-family of very high excitation OPARGs at the top of the AGN wing. This group is slightly displaced towards the left of the rest of the AGN galaxies, suggesting a stronger ionizing radiation field with respect to the gas pressure. 
}
{
Only  very-high excitation radio galaxies (VHERGs) have Eddington ratios higher than $10^{-2}$, which are canonically considered as the lower limit for the occurrence of radiative efficient accretion. If our estimates of the bolometric luminosities are correct, this means than only a small proportion of mainstream HERGs are indeed radiatively efficient.
}

\keywords{Surveys -- Galaxies: active -- Radio continuum: galaxies
}

\maketitle


\section{Introduction}
\label{intro}

The first systematic studies of radio galaxies focussed on the diversity of radio morphologies. \citet{Fanaroff.Riley.1974a} defined two morphological classes of extended radio galaxies. Class I (FR I) consisted of objects where the low-brightness regions are located further away from the galaxy, and Class II (FR II) in which it is the high-brightness regions that are located farther from the galaxy. Later, systematic spectroscopic studies of radio galaxies found that the spectra can also be divided into two classes \citep{Hine.Longair.1979a}, class A showing strong emission lines such as Seyfert galaxies and class B without strong emission lines. There is no one-to-one relation between these morphological and spectroscopic classes. While most FR II objects show strong high-ionization emission lines, both FR I and FR II objects are found among objects without strong emission lines.

Further studies suggested that the spectra of radio galaxies hold potential information on the accretion mode on the central black hole. In contrast to objects with strong emission lines, later dubbed high-excitation radio galaxies (HERGs) \citep{Laing.etal.1994a, Chiaberge.Capetti.Celotti.2002a, Hardcastle.Evans.Croston.2007a}, objects without strong lines  --  low-excitation radio galaxies (LERGs) -- are considered to have a low accretion efficiency, leading to the entire accretion energy being channeled into jets. 
\citet{Hardcastle.Evans.Croston.2007a} suggested that LERGs are powered by accretion of the hot phase of the intergalactic medium, while HERGs, experimenting a radiative efficient accretion, require fueling from cold gas found in the vicinity of accretion disks, a view adopted by \citet{Kauffmann.Heckman.Best.2008a} and \citet{Heckman.Best.2014a}, for example. However, \citet{Hardcastle.2018b} argued that radiative efficiency in radio galaxies is governed by the  accretion rate on to the massive black hole and not by the type of accreted matter (\ie hot or cold).

While there is a consensus on the existence of two modes (the radiative mode or quasar mode which is believed to occur in HERGs as opposed to the jet mode or radio mode which is believed to occur in LERGs), the practical distinction between the two categories has changed over the years. \citet{Laing.etal.1994a} defined HERGs as objects with $\oiii/\Ha  > 0.2$ and $\woiii > 3$~\AA, where $\woiii$ is the equivalent width of the \Oiii emission line.  \citet{Jackson.Rawlings.1997a} defined LERGs as objects with a \woiii smaller than 10~\AA\ or $\oii/\oiii >1$ (or both).  \citet{Chiaberge.Capetti.Celotti.2002a} defined HERGs as objects with a \woiii larger than 10 \AA\ and \oiii/\oiii $ >0.1$.  Based on the spectroscopy of 113 CR sources with redshifts smaller than 0.3, \citet{Buttiglione.etal.2010a} constructed an excitation index ($EI$) defined as $\log \oiii/\Hb - 1/3 (\log \nii/\Ha + \log \sii/\Ha + \log \oi/\Ha)$. They found this parameter to be bimodal and placed the limit for LERGs at $EI < 0.95$.  \citet{Best.Heckman.2012a} classified their 7302 radio active galactic nuclei (AGNs) from the Sloan Digital Sky Survey (SDSS) main galaxy sample \citep{Strauss.etal.2002a} in the redshift range $0.01 < z < 0.3$. Since the Buttiglione index could not be computed for many of these objects, they applied a complex approach to distinguish HERGs and LERGs based on a combination of available line ratios.  \cite{Pracy.etal.2016a} came back to a simpler classification, considering as HERGs those galaxies with a signal-to-noise ratio (S/N) at 5007~\AA\ larger than 3 and $\woiii > 5$~\AA, a view also adopted by \citet{Prescott.etal.2016a} and \citet{Ching.etal.2017a}.

As a matter of fact, none of these definitions were based on a physical criterion. Except for the definition by \citeauthor{Buttiglione.etal.2010a}, which unfortunately is not usable for all objects, they are not even justified by an observational dichotomy.
In this paper, we propose a simple and physical criterion to separate radio galaxies into two groups:  optically active radio galaxies (OPARGs) and optically inactive radio galaxies (OPIRGs).

Our sample of radio galaxies is based on the ROGUE~I and ROGUE~II catalogues (\citealp{KozielWierzbowska.Goyal.Zywucka.2020a} and 2024 in preparation), which are the largest human-made catalogues of radio sources associated with optical galaxies. The galaxies are from SDSS (\citealp{York.etal.2000a}), which allowed us to obtain redshifts and investigate such properties as radio luminosities and morphologies, galaxy masses and ages,  emission-line properties, among others.

The paper is organized as follows. In Sect.~\ref{data} we present the radio and optical data that were used for this study. In Sect.~\ref{def} we explain how we separated our sample into two populations and why.  In Sect.~\ref{populations}, in a way similar to what has been done for HERGs versus LERGs, we compare the  populations of OPARGs and OPIRGs. In Sect.~\ref{general} we consider the OPARGs and OPIRGs in the general context of SDSS galaxies. In Sect.~\ref{BPT} we placed the OPARGs in the \citet[BPT]{Baldwin.Phillips.Terlevich.1981a} diagram and discuss how their positions differ from those of the bulk of SDSS galaxies with AGN. In Sect.~\ref{Eddington} we estimate the radiative efficiencies of the  OPARGs and in Sect.~\ref{Eddaccretion} we discuss the Eddington-scaled accretion rates of OPARGs and OPIRGS. In Sect. \ref{theory}, based on these observationally derived parameters, we  investigate  the accretion properties and jet launching mechanisms in the different families of AGN-containing galaxies.   
Sect.~\ref{sum} presents a summary of our main findings. In an appendix we propose a method to derive the AGN bolometric luminosities, taking into account the contribution of star formation to the emission lines observed in AGN galaxies. These bolometric luminosities were used to derive the Eddington ratios.

\section{Data} 
\label{data}


\subsection{The original sample of radio sources in the ROGUE catalogues}
\label{radiosources}

The parent samples of galaxies in the ROGUE catalogues are the main galaxy sample (MGS, \citealp{Strauss.etal.2002a}) and the Red Galaxy Sample (RGS, \citealp{Eisenstein.etal.2001a}) of the SDSS DR7 \citep{York.etal.2000a, Abazajian.etal.2009a}.  Only those galaxies whose optical spectra are good enough (S/N in the continuum at 4020~\AA\ of at least 10) to allow the study of stellar population and emission line properties were retained, resulting in a sample of 673,807 galaxies. For each of these galaxies, radio counterparts were searched in the FIRST \citep{White.etal.1997a} and NVSS \citep{Condon.etal.1998a} radio catalogues. ROGUE I contains those radio sources for which FIRST data present a core and counts 32,616 objects, while ROGUE II contains coreless radio sources and counts 7,694 objects. The ROGUE catalogues contain information on radio fluxes, radio morphologies as well as the morphologies of the parent galaxies. 
 
In this paper, we retained only those sources present in the Main Galaxy Sample, which is a complete sample of galaxies (excluding Type I AGNs) down to $r$-band Petrosian magnitudes of 17.77.
Throughout this work we consider a cosmology with
$H_0 = 70 \ {\rm km} \ {\rm s}^{-1} \ {\rm Mpc}^{-1}$, $\Omega_M=0.30$,
and $\Omega_{\Lambda}=0.70$.

\subsection{SDSS products}
\label{sdss}

Each SDSS spectrum has been processed with the inverse stellar population synthesis code \starlight \citep{CidFernandes.etal.2005a}.  This code recovers the stellar population of a galaxy by fitting a pixel-by-pixel model to the observed spectrum (excluding narrow windows where emission lines are expected). The model is a linear combination of 150 `simple stellar populations’ templates of 25 ages between 1 Myr and 18 Gyr and  6 metallicities between 0.005 and 2.5 Z$_\odot$ using \citet{Bruzual.Charlot.2003a} evolutionary stellar population models.
The stellar dust attenuation $A_V$ is obtained adopting a \citet{Cardelli.Clayton.Mathis.1989a} extinction law with $R_V = 3.1$ by requiring that the reddened modelled spectrum matches the observed one.
The intensities of the emission lines were measured by Gaussian fitting after subtracting the modelled stellar spectrum from the observed one. More details can be found in \citet{CidFernandes.etal.2005a}.
The total stellar masses of the galaxies, \Mstar, were obtained as in \citet{CidFernandes.etal.2005a}, assuming that the mass-to-light ratios are the same outside and inside the fibre and scaling the stellar masses encompassed by the fibre by the ratio between total (from the photometric database) and fibre $z$-band luminosities.  We also make use of some parameters related to the star-formation histories and stellar mass growth extracted from the \starlight database \citep{CidFernandes.etal.2009a}.
The black-hole mass, \MBH, is estimated from the stellar velocity dispersion determined by \starlight, $\sigma_\star$, using the relation by \citet{Tremaine.etal.2002a}.
The mean stellar age weighted by luminosity ($t_L$) is obtained as explained in \citet{CidFernandes.etal.2005a}.

\subsection{Radio galaxies from ROGUE I and ROGUE II}
\label{radiogalrogue} 

\begin{figure} 
  \centering
  \includegraphics[width=0.8\columnwidth, trim=10 15 10 10, clip]{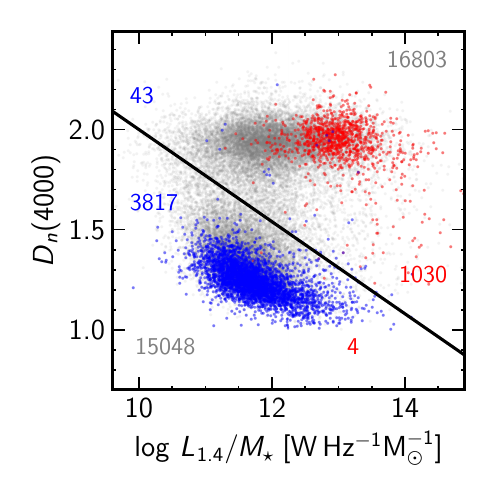}
  \caption{DLM diagram for ROGUE radio sources from the MGSz sample. Blue points are pure SF galaxies as defined by  \citet{Stasinska.etal.2006a} on the \Nii/\Ha versus \Oiii/\Hb plane.
 Red points are the secure FR I and FR II radio sources from our sample.   }
  \label{fig:DLM}
\end{figure} 

The ROGUE catalogues do not give any clue on the origin of the radio emission of their sources. To distinguish sources where the radio emission is due to a radio AGN from those where it is powered by processes linked to star formation we use the $D_n(4000)$ versus $L_{1.4}/$\Mstar (DLM) diagram,  where  $D_n(4000)$ is the discontinuity at 4000 \AA\ taken from the synthetic spectra as explained in \citet{Stasinska.etal.2006a} and $L_{1.4}/$\Mstar is the ratio between the total luminosity at 1.4 GHz and the stellar mass obtained by \starlight. As shown by \citet{KozielWierzbowska.etal.2021a} the DLM digram is a very efficient stand-alone diagram to distinguish these two families of radio sources. We note that this diagram has been used by many authors in the past \citep[\eg][]{Best.etal.2005a, Best.Heckman.2012a, Sabater.etal.2019a}, but always in conjunction with other diagrams. As argued by \citet{KozielWierzbowska.etal.2021a}, we prefer using just one criterion, providing a simpler classification and an easier discussion of selection effects.

Figure~\ref{fig:DLM} shows the DLM diagram for the ROGUE objects with redshifts between 0.002 and  0.4 in the MGS sample. The lower redshift limit  makes sure that luminosity distances are not dominated by peculiar motions (\eg \citealp{Ekholm.etal.2001a}), while the upper redshift limit ensures that the \Ha\ line lies in the observable wavelength range. In the following, MGSz will refer to the MGS sample with these redshift cuts and the quality cuts defined in Sect.~\ref{radiosources}. and MGSz$^\prime$ the MGSz sample excluding AGN radio galaxies according to the DLM.
Red symbols are FR I and FR II radio sources, classified by visual inspection of radio maps in the ROGUE I and II catalogues \citep{KozielWierzbowska.Goyal.Zywucka.2020a}, that is to say objects which are undoubtedly radio AGNs. Blue symbols are objects whose position in the BPT diagram (\oiii/\Hb versus \nii/\Ha with a minimum S/N of 3 in each line; see diagrams in Sect.~\ref{BPT}) indicate that they are pure star-forming (SF) galaxies according to the \citet{Stasinska.etal.2006a} criterion.

Grey points are the remaining objects of our sample. The black line is the dividing line proposed by \citet{KozielWierzbowska.etal.2021a} to separate radio-AGNs from other radio sources. The total number of galaxies on each side of the dividing line is indicated in grey. The numbers of pure SF galaxies on each side of the dividing line are in blue. The numbers  of FR I or FR II  galaxies on each side of the dividing line are in red. Among 1034 FR I or FR II galaxies, we can see that only 4 are misclassified as being star-forming. In total, our sample of radio-galaxies consists of 16\,803 objects.

\begin{figure*} [!ht]
  \centering
  \includegraphics[width=1\linewidth, trim=0 30 0 0]{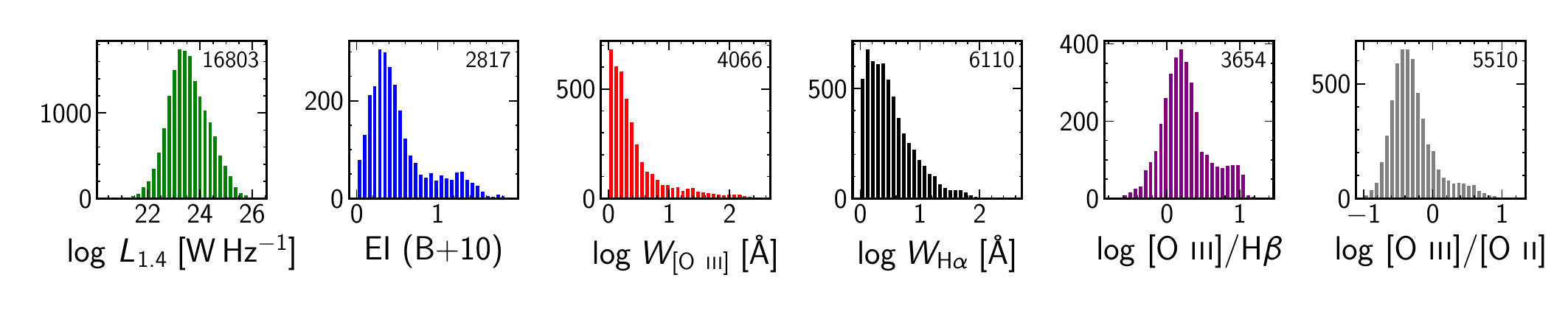}
  \caption{Histograms of several properties of 
  ROGUE AGN radio sources: \Lrad, the excitation index as defined by \citet{Buttiglione.etal.2010a}, $\log \woiii$, $\log \wha$, $\log\, \oiii/\Hb$, and $\log\, \oiii/\oii$. The number of objects for which these properties can be defined is indicated at the top right (see text).}
  \label{fig:histos}
\end{figure*}

\section{Defining two spectral families of radio galaxies} 
\label{def}

\subsection{Searching for dichotomies}
\label{dichotomies}

One obvious first step is to search whether there is a clear dichotomy in some properties of the objects to classify. Figure~\ref{fig:histos} shows histograms of the values of parameters that come to mind: the total radio luminosities \Lrad, the excitation index of \citet{Buttiglione.etal.2010a}, the equivalent width of \oiii, \woiii, the equivalent width of \Ha, \wha, the \oiii/\Hb ratio and the \oiii/\oii ratio. For the histograms involving line ratios, a S/N cut of $3$ was adopted for each line. In each histogram we indicate the number of objects that has a value for the corresponding parameter within the range shown in the plots. The total number of objects in our sample that can be classified using \woiii or \wha is of course equal to 16\,803, and not to the values indicated on the corresponding panels, which only refer to positive equivalent widths.

The histogram of radio luminosities is very smooth with a conspicuous maximum and does not reveal any dichotomy at all. The clearer cases of a dichotomy are those of the \citet{Buttiglione.etal.2010a} index and of \oiii/\Hb. However, only a small proportion of objects can be classified in this way, due to S/N issues. The \oiii/\oii diagram shows a tail at log \oiii/\oii larger than about 0.2. However, this parameter also cannot be computed for all the objects. As regards \woiii and \wha, the observed values are reliable only when they are larger than 1~\AA, smaller values are equivalent to zero (and they represent the majority of the objects). However, \woiii as well as \wha can be used as classifiers, since all the objects can be assigned a value, either zero or positive. The most commonly adopted classification of radio galaxies  uses a value of $\woiii = 5$~\AA\ to distinguish between HERGs and LERGs. However we do not see any dichotomy so  5~\AA\ seems an arbitrary value. We do see an extended tail at log \woiii larger than 1, and the objects it concerns are likely the same as the ones with high values of \oiii/\Hb and \oiii/\oii. These objects deserve more attention (see Sect.~\ref{BPT}). The  \wha histogram  is similar to the \woiii histogram, but with a less pronounced tail at $\log \wha \gtrsim 1.3$. Here again no dichotomy is seen, but we will argue in the next section why \wha is a good parameter to classify radio galaxies.

\begin{figure} [!ht]
  \centering
  \includegraphics[width=0.65\columnwidth, trim=0 10 0 10]{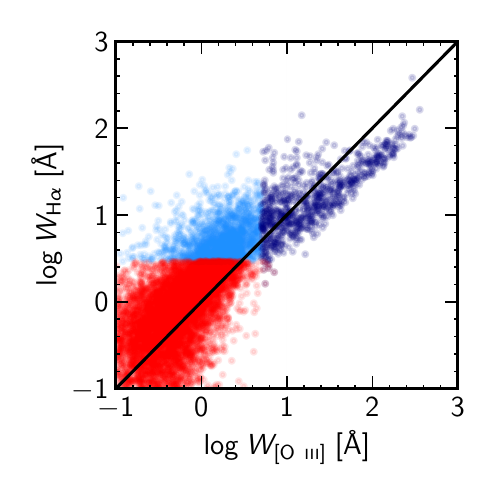}
  \caption{Relation between $\log \woiii$ and $\log \wha$. Objets with \wha smaller than 3 \AA\ are in red, those with \woiii larger than 5 \AA\ are in dark blue. }
  \label{fig:WHaWO3}
\end{figure}
\subsection{Our proposition}
\label{proposition}

We want to use a simple criterion that can be equally applied to all the objects of the sample. This is indeed the trend adopted by most recent studies, which use \woiii as the main criterion. This justifies the nomenclature HERGs and LERGs for high-and low-excitation radio galaxies, since the definition of an object `excitation' is linked to the strength of lines from highly ionized elements (here \Opp). 

However, it is not clear why it was the \oiii line which was first adopted to make the distinction between two classes of galaxies. May be because it was the most prominent emission line in optical spectra. In later studies, the two families of radio galaxies were interpreted in terms of Eddington ratio \citep{Kauffmann.Heckman.2009a, Best.Heckman.2012a} which was computed using the luminosity of the \oiii\ line. However, as already pointed out in \citet{Netzer.2009a}, \citet{KozielWierzbowska.Stasinska.2011a} and \citet{Sikora.etal.2013a}, the bolometric luminosity of AGNs should \textit{a priori} be  more directly linked to \LHa than to \Loiii, even if \Ha\ is not the strongest optical line in the spectrum. There are two reasons for this. First,  from recombination theory the  intensity of  \Ha\ is directly proportional  to that of the  hydrogen Lyman $\alpha$ line, which is 8 times stronger and is generally the most intense line emitted by a photoionized object. This argues for  \Ha\  being a good  indicator of the total energy emitted in the lines. Second, \Ha\  measures exactly the total number of ionizing photons absorbed by the gas, irrespective of the nebula excitation (see however Sec. \ref{boloparg})\footnote{For type II AGNs in SDSS, the \Ha and $\nii\lambda\lambda6548,6584$ lines may overlap, but their peaks are typically well separated. The deblending is handled by fitting simple Gaussian profiles to the \Ha and \nii lines.}.

Another, completely different reason to use \Ha, is that the presence of hot low-mass evolved stars (HOLMES) in retired galaxies (galaxies which have stopped forming stars; \citealp{Stasinska.etal.2008a}) can also produce weak emission lines not at all related to the AGN phenomenon. From a stellar population analysis of SDSS galaxies, \citet{CidFernandes.etal.2011a} showed that in galaxies having emission lines with $\wha < 3$~\AA\ the ionization is produced by HOLMES and not by weak AGNs (previously called LINERs for low-ionization nuclear emission-line regions, \citealp{Heckman.1980a}).
In  Appendix \ref{MANGA}, we use integrated field unit observations from the Mapping Nearby Galaxies at Apache Point Observatory (MaNGA) project \citep{Bundy.etal.2015a} to estimate the proportion of galaxies with $\wha < 3$~\AA\ that host a very weak AGN. We find that it is very small.

Therefore, we decided to divide our sample of radio galaxies in two classes as follows: `optically inactive radio galaxies' (OPIRGs)\footnote{Our OPIRGs remind XBONGs (X-ray bright, optically normal galaxies) or `optically dull AGNs' discovered from X-ray surveys \citep[\eg][]{Trump.etal.2009a, Smith.Koss.Mushotzky.2014a, Agostino.etal.2023a}.}   where either no emission lines are seen at all or where  $\wha < 3$~\AA, and `optically active radio galaxies' (OPARGs) where $\wha \ge 3$~\AA. While this division reminds of the LERG/HERG classification, it has nothing to do a priori with excitation.  

Figure~\ref{fig:WHaWO3} shows \wha\ versus \woiii\ for our sample of radio galaxies. OPIRGs are represented by red points. Objects with \woiii $>$ 5 \AA\ are represented by dark blue points. In our sample, there are 14082 OPIRGs and 2721 OPARGs. Note that the number of HERGs by the condition $\woiii > 5$~\AA\ (the points in dark blue) would have been only of 804 in our sample.

\subsection{Relation with excitation}
\label{excitation}
 
\begin{figure} 
  \centering
  \includegraphics[width=1\columnwidth, trim=10 30 10 0]{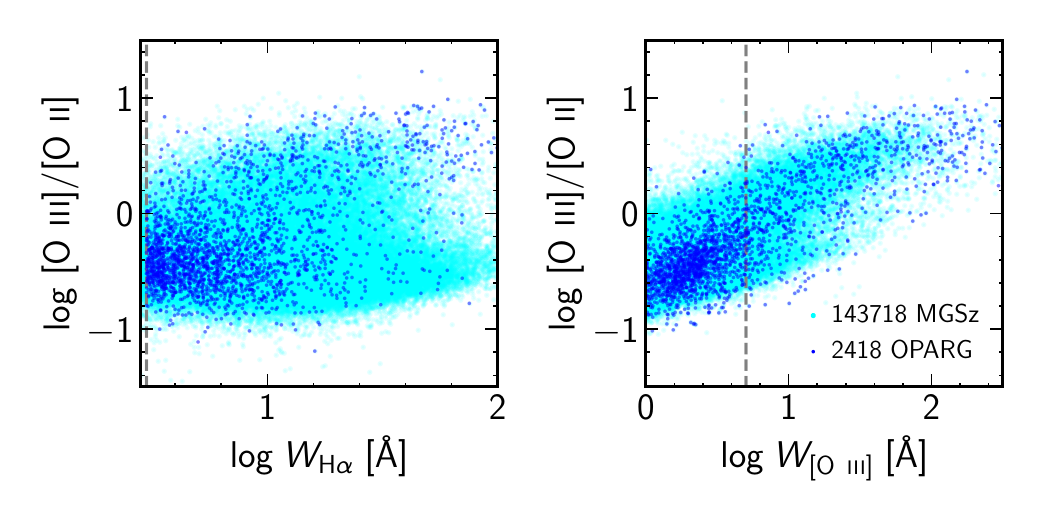}
  \caption{Excitation ratio \oiii/\oii as a function of \wha (left) and \woiii (right). Only objects with a S/N of at least of 3 have been plotted. Dark blue is for OPARGs. Cyan is for `active' galaxies from the MGSz sample according to the \citet{Stasinska.etal.2006a} line (see text).}
  \label{fig:WHaWO3O3O2}
\end{figure}

Figure~\ref{fig:WHaWO3O3O2} shows the values of \oiii/\oii, which is a proper measure of excitation, as a function of \wha (left) and \woiii (right). The dark blue points are our OPARGs. The points in cyan represent optically active galaxies from the MGSz sample according to the \citet{Stasinska.etal.2006a} dividing line in the \oiii/\Hb versus \nii/\Ha plane (with a minimum S/N of 3 in those emission lines). From the right panel, we see that, not surprisingly, \woiii is strongly correlated with \oiii/\oii, so that \woiii is indeed linked to excitation. On the other hand, the \oiii/\oii versus \wha diagram shows two branches, one of higher excitation, the other of lower excitation. The proportion of high-excitation radio galaxies at a given \wha is larger than that of active galaxies that do not emit in radio. We also note that the radio galaxies reach higher equivalent widths than active galaxies that do not emit in radio.  We will come back to this later.

\begin{figure*} 
  \centering
  \includegraphics[width=0.77\linewidth, trim=0 30 0 0]{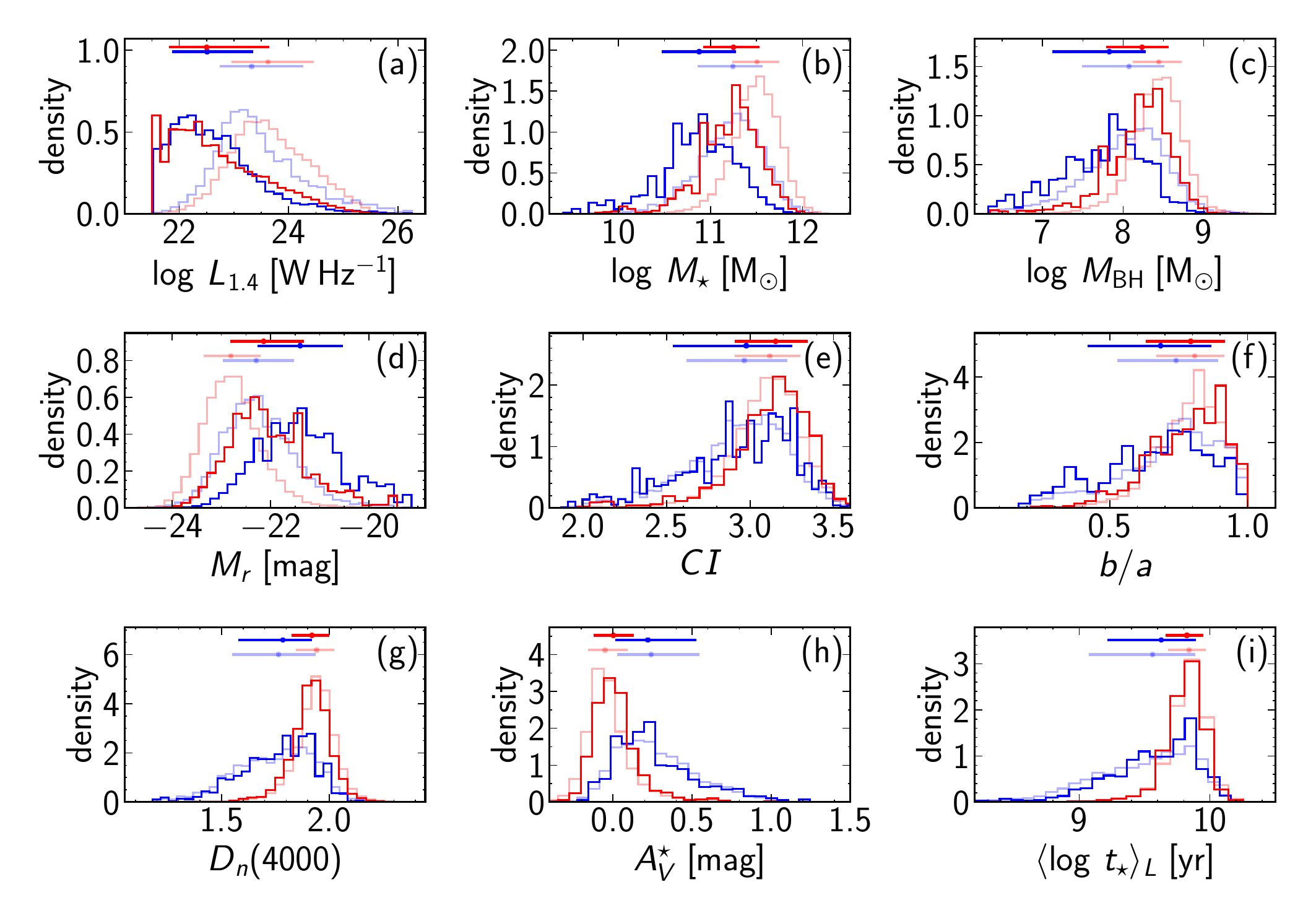}
  \caption{Histograms of the distributions of various parameters in the OPARG and OPIRG samples:  the radio luminosity \Lrad, the total stellar mass \Mstar, the black hole mass \MBH, the galaxy concentration index $CI$, the galaxy flatness parameter $b/a$, the stellar extinction $A_V$, and the mean stellar age, $t_L$.
 OPARGs are shown in blue and OPIRGs in red; darker lines represent density histograms with the completeness correction, and lighter lines represent density histograms without it. The segments at the top of each panel show the 16 to 84 percentiles of each distribution, with the dot marking the median. }
  \label{fig:Vmax}
\end{figure*}

\begin{figure*} 
  \centering
  \includegraphics[width=1\linewidth, trim=40 25 40 0]{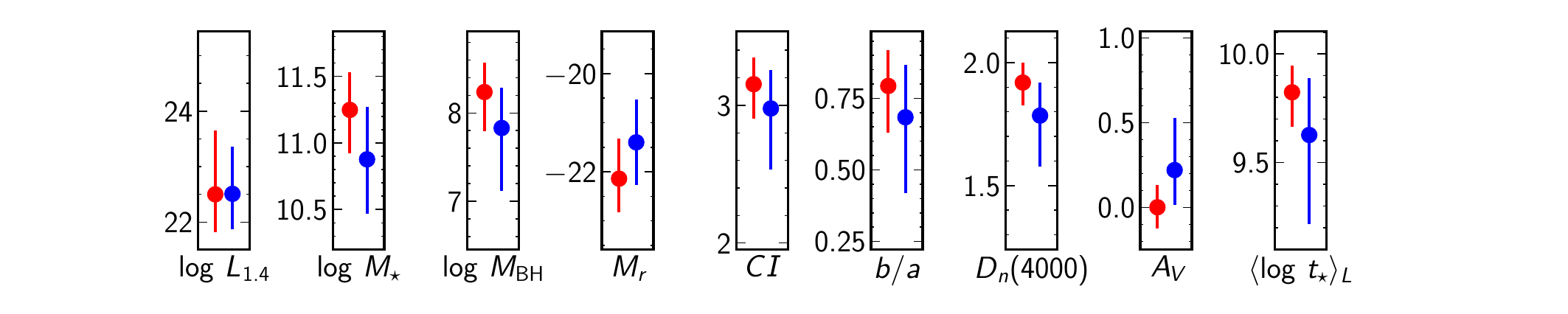}
  \caption{Median, 16, and 84 percentiles of the distributions (corrected for completeness) of OPARGs and OPIRGs, for the same parameters as in Fig.~\ref{fig:compare-OPARG-OPIRG}. OPARGS are in blue, and OPIRGs are in red.}
  \label{fig:compare-OPARG-OPIRG}
\end{figure*}

\begin{figure*} 
  \centering
  \includegraphics[width=1.\linewidth, trim=40 25 40 0]{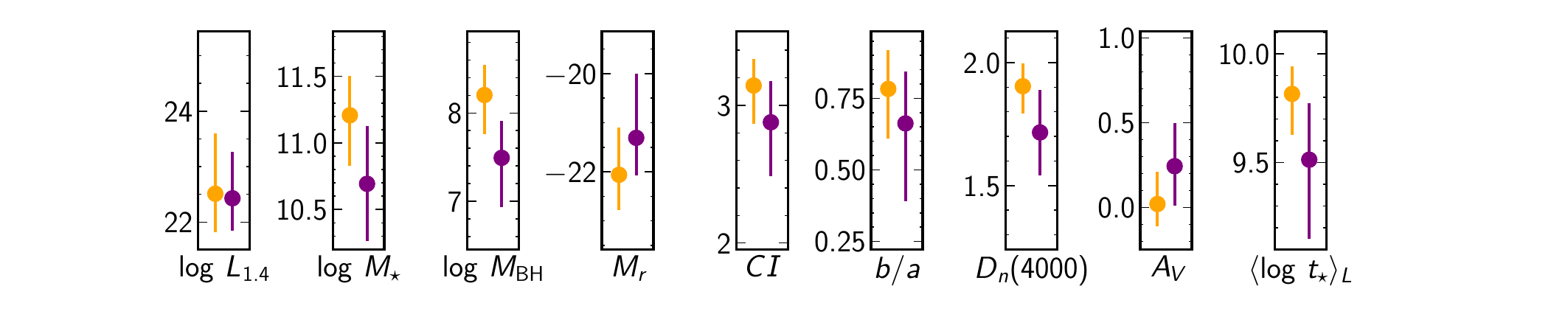}
  \caption{Median, 16, and 84 percentiles of the distributions (corrected for completeness) of HERGs and LERGs, for the same parameters as in Fig.~\ref{fig:compare-OPARG-OPIRG}. HERGs are in purple, and LERGs are in orange.}
  \label{fig:compare-HERG-LERG}
\end{figure*}

\section{The populations of OPARGs and OPIRGs}
\label{populations}

 It is customary to compare the main characteristics of HERGs and LERGs (\eg \citealp{Heckman.Best.2014a}). Here we will compare the properties of OPARGs and OPIRGs. For a meaningful comparison, one has first to take into account the fact that the limiting magnitudes of the MGS are $14.5 \le m_r \le 17.77$ \citep{Strauss.etal.2002a} and the radio flux limit is 1~mJy \citep{KozielWierzbowska.Goyal.Zywucka.2020a}. If the two populations differ in luminosities the comparison results may be biased. Therefore, it is first necessary to apply a completeness correction. 

\subsection{Completeness correction}
\label{incompleteness}

For each galaxy we define the comoving volume $V_\mathrm{max}(m_r, F_{1.4})$ in which a galaxy can be observed in both the optical and radio catalogues as $V_\mathrm{max} = V_\mathrm{out} - V_\mathrm{in}$. $V_\mathrm{out}$ is the outermost volume in which a galaxy may be observed given both the optical magnitude and the radio flux limits, whereas $V_\mathrm{in}$ is the innermost volume in which a galaxy may be observed due to the brightest optical magnitude limit. Physical properties are thus compared in terms of their volume densities, that is, weighted by $1/V_\mathrm{max}$ \citep{Schmidt.1968a, Condon.1989a}.

\subsection{Comparing the two populations}
\label{compare}

Figure~\ref{fig:Vmax} shows the histograms of the distributions of various parameters of interest. In the first row we have \Lrad, \Mstar and \MBH. In the second row we have parameters linked to the optical properties of our galaxies: The absolute magnitude in the $r$ band, $M_r$, the galaxy concentration index, $CI$, and the major-to-minor axis ratio, $b/a$, being equal to 1 for a spherical object. The third row shows the values of the 4000-\AA\ break $D_n(4000)$, the dust attenuation $A_V$ and the log of the mean stellar age $t_\star$. All the parameters except the radio luminosity have been derived from the \starlight analysis of the galaxy spectra. OPARGs are shown in blue and OPIRGs in red; darker and lighter lines represent density histograms with and without the completeness correction respectively. The horizontal segments at the top show the 25 to 75 percentile range of each distribution, with the dot marking the median.

Figure~\ref{fig:compare-OPARG-OPIRG} represents the results of Fig.~\ref{fig:Vmax} with the completeness correction in a compact form.
It compares properties of OPARGs and OPIRGs by plotting their 16, 50 and 84 percentiles. OPARGs are represented in blue, OPIRGs in red. The properties that are reported are the same as in Fig. \ref{fig:Vmax}.
Figure~\ref{fig:compare-HERG-LERG} is the same as Fig.~\ref{fig:compare-OPARG-OPIRG} but for HERGs and LERGs using the $\woiii = 5$~\AA\ criterion.

The absolute magnitudes $M_r$ indicate that, on average, OPIRGs have higher optical luminosities than OPARGs, which justifies the use of $V_\mathrm{max}$ to compare the two populations.

Surprisingly, the radio luminosity distributions of OPARGs and OPIRGs are very similar. Actually, the same happens between HERGs and LERGs using the definitions of \citet{Laing.etal.1994a} and \citet{Pracy.etal.2016a}. This is at odds with the view of \citet{Heckman.Best.2014a} who state in their Fig.~4 that high excitation radio galaxies have high radio luminosities while low-excitation radio galaxies have moderate radio luminosities. 

On the other hand, the stellar masses and black hole masses are larger for OPIRGs than for OPARGs, and larger for LERGs than for HERGs, in agreement with the common view. 

Significant differences are also found for the spectral discontinuity, $D_n(4000)$ (which decreases with increasing mean stellar age) and for the mean stellar age: OPARGs contain young stellar populations, while the mean stellar age of OPIRGs lies in a narrow range between $10^{9.75}$ and $10^{9.92}$ yr. 
A large difference between OPARGs and OPIRGs (and between HERGs and LERGs) is also seen in the stellar extinction, which is much larger in OPARGs, indicating the presence of dust (and gas). Thus, while OPIRGs are galaxies without any sign of recent star formation, OPARGs may be presently forming stars, or at least star formation has stopped recently. This is well in agreement with the finding of other groups \citep{Janssen.etal.2012a, Best.Heckman.2012a} using different approaches.

\begin{figure} 
  \centering
  \includegraphics[width=1\linewidth, trim=10 30 20 0]{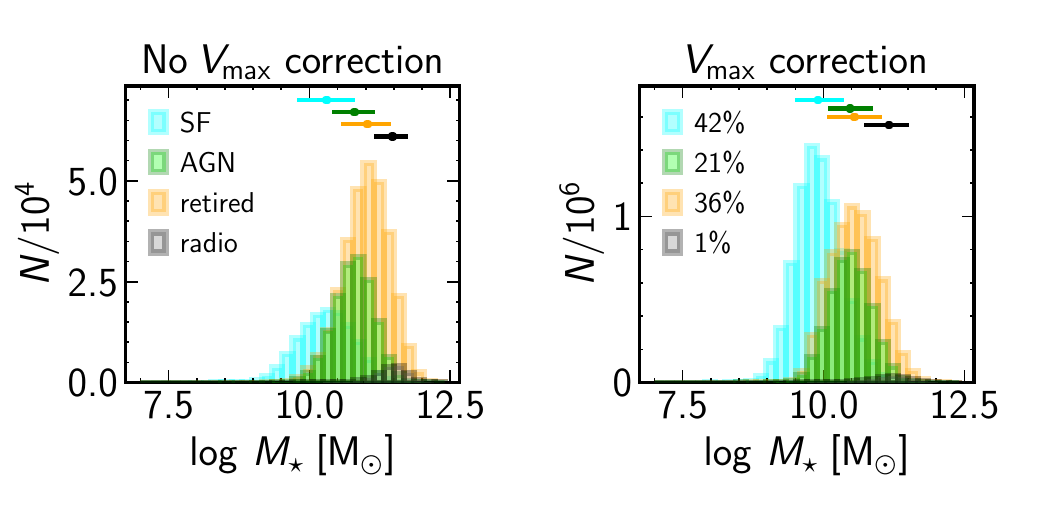}
  \caption{Histograms of the distributions of SF galaxies (light blue), AGN hosts (green), retired galaxies (orange), and radio galaxies (black) in the MGSz sample. The horizontal segments show the 16 to 84 percentiles of each distribution, with the dot marking the median. Left: without $V_\mathrm{max}$ correction; right: with $V_\mathrm{max}$ correction. Percentages show the proportion of each subsample after applying the completeness correction.
    }
  \label{fig:histo-sf-agn-ret-radio}
\end{figure}

\begin{figure} 
  \centering
  \includegraphics[width=1\linewidth, trim=20 30 20 0]{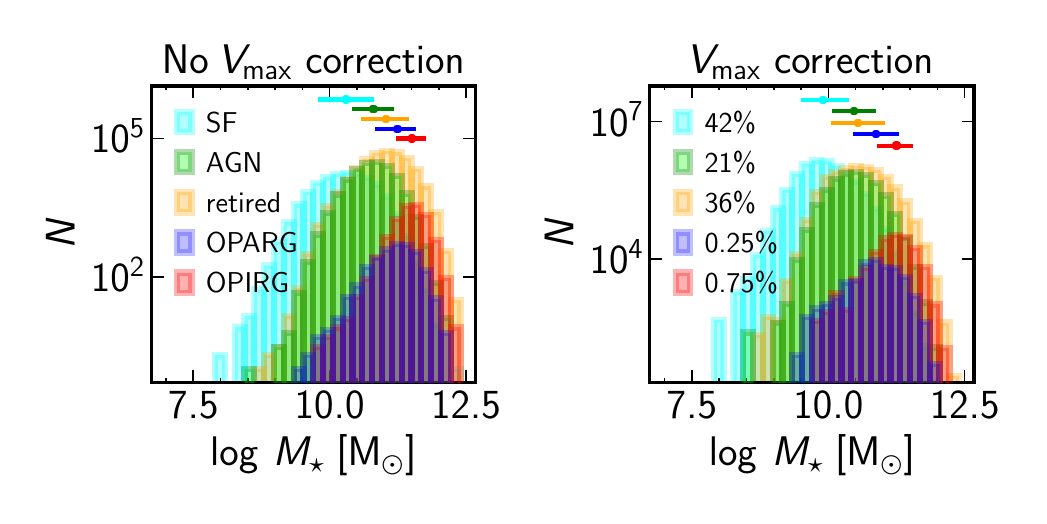}
  \caption{Same as Fig.~\ref{fig:histo-sf-agn-ret-radio} but with ordinates in logarithmic scale. The sample of radio galaxies has been divided into OPIRGs (red) and OPARGs (dark blue).}
  \label{fig:hist-all}
\end{figure}

\section{OPARGs and OPIRGs in the general context of SDSS galaxies}
\label{general}

Since we have a complete sample of radio galaxies extracted from a complete sample of SDSS galaxies, it is interesting to put everything in context. 
Figure~\ref{fig:histo-sf-agn-ret-radio} show a histogram of the various types of galaxies that form the MGSz sample. 
The left panel shows the histogram of observed galaxies, and the right panel shows the histograms after the correction for completeness explained in Sect.~\ref{incompleteness}. The galaxies have been divided into categories using the WHAN diagram \citep{CidFernandes.etal.2011a} in the simplified following way:
\vspace{-0.7em}
\begin{itemize}
  \item Pure SF galaxies: $\log \nii/\Ha\ \le -0.4$ and  $W_{\Ha} > 3$ \AA;
  \item AGN hosts: $\log \nii/\Ha\ > -0.4$ and  $W_{\Ha} > 3$ \AA;
  \item Retired galaxies: $W_{\Ha} \le 3$ \AA,
\end{itemize}
\vspace{-0.5em}
retired galaxies being galaxies that stopped forming stars but may have emission lines excited by HOLMES.

We can see that when the histograms are corrected for incompleteness,  the proportion of SF galaxies is much larger, due to the fact that, being less massive they are also less luminous in the $r$ band. However, the figure ignores many dwarf galaxies, especially dwarf elliptical galaxies, which are too faint to be detected by the SDSS. 
The figure shows also the well-known fact that SF galaxies are on average less massive than galaxies with an optically active AGN which in turn are less massive, on average, than retired galaxies \citep{CidFernandes.etal.2011a}. However, there is an important overlap between these families, especially between galaxies with an optically active AGN and retired galaxies. 
Radio galaxies are only found among the most massive galaxies, with a median stellar mass of $\log \Mstar/\mathrm{M_\odot} = 11.47$ and 16 and 84 percentiles at $11.18$ and $11.72$ (or, with the completeness correction, a median of $\log \Mstar/\mathrm{M_\odot} = 11.17$ and 16 and 84 percentiles
at $10.75$ and $11.48$). In total, they represent 1\% of the total population of galaxies after correction for the Malmquist bias.

Figure~\ref{fig:hist-all} is similar to Fig.~\ref{fig:histo-sf-agn-ret-radio} but the ordinates are now in logarithm, so that one can see the details of the population of radio galaxies, which represent only a small portion of a total sample of galaxies. 
This figure shows that OPIRGs cover the upper stellar-mass range of retired galaxies, while OPARGs cover the upper stellar-mass range of galaxies hosting optical AGNs. The percentages after completeness correction show that OPIRGs are three times more abundant than OPARGs.


\begin{figure} 
  \centering
  \includegraphics[width=1\linewidth, trim=10 30 10 0]{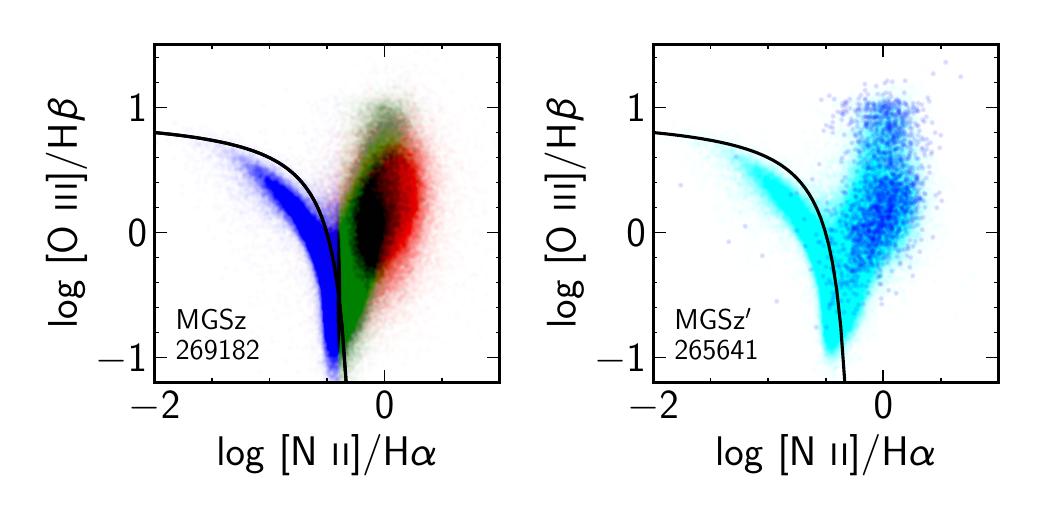}
  \caption{BPT diagram for all the MGSz galaxies with an S/N of at least 3 in all the relevant lines. Left: pure SF galaxies are in blue, galaxies containing an AGN are in green; retired galaxies are in orange. Right: blue points represent OPARGs, superimposed on the remaining MGSz$^\prime$ galaxies (in cyan).}
  \label{fig:BPThergSN3.png}
\end{figure}

\begin{figure} 
  \centering
  \includegraphics[width=1\linewidth, trim=10 25 0 0]{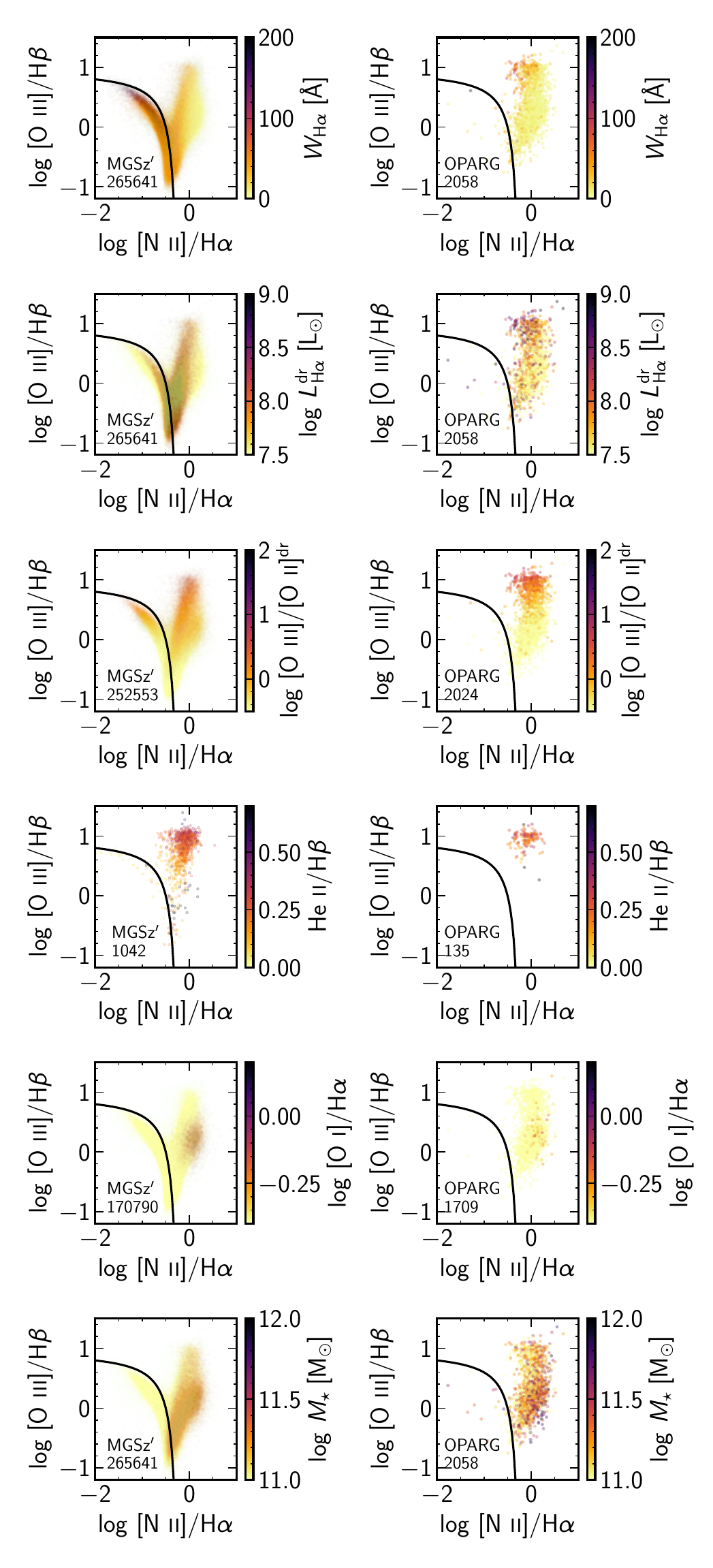}
  \caption{BPT diagram for the whole MGSz$^\prime$ sample (left) and for OPARGs (right). Colour-coding is done with respect to specific parameters. From top to bottom: \wha, \LHa (corrected for extinction), \oiii/\oii (dereddened), \heii/\Hb, \oi/\Ha, and \Mstar.}
  \label{fig:BPT2}
\end{figure}

\section{OPARGs in the BPT diagram} 
\label{BPT} 

The  \oiii/\Hb vs \nii/\Ha diagram,  so-called BPT after \citet{Baldwin.Phillips.Terlevich.1981a} is a very popular diagnostic diagram allowing to distinguish pure SF galaxies from the rest (galaxies containing AGNs and retired galaxies).\footnote{It was originally used to distinguish SF galaxies from galaxies containing AGNs and `LINERs', but as shown by \citet{Stasinska.etal.2008a, Yan.Blanton.2012a} not all the galaxies in the right branch contain an AGN. }
The left panel of Figure  \ref{fig:BPThergSN3.png} shows all the MGSz galaxies (including the radio galaxies) within redshifts 0.002 and 0.4  and signal-to-noise (S/N) ratio larger than 3 in all the relevant lines. Blue points represent pure SF galaxies, green points represent galaxies containing an AGN, and red points represent retired galaxies as defined in Sect. \ref{general}. In the following, for convenience, we will refer to these  galaxies containing an AGN as Seyferts (including `true LINERs' and radio galaxies).

The right panel of Figure  \ref{fig:BPThergSN3.png} shows the BPT diagram with OPARGs having S/N larger than 3 in all the relevant lines as blue points, superimposed on all the galaxies from the MGSz$^\prime$ sample (\ie all the MGSz excluding the AGN radio galaxies) with same S/N limitations in cyan.
In both panels, the curve in black is the  line proposed by \cite{Stasinska.etal.2006a} to delimitate pure SF galaxies from the rest.
Two families of OPARGs can be distinguished, the higher excitation one being slightly dislocated to the left with respect to the main stream of SDSS galaxies containing an AGN.


In order to get some clue of the properties of these high excitation OPARGs, in Fig. \ref{fig:BPT2} we plot again the galaxies in the BPT diagram, but now colour-coding the points according to the values of various parameters. From top to bottom: 
\wha, $\log \LHa$ corrected for extinction, $\log\, \oiii/\oii$ corrected for reddening, \Heii/\Hb, $\log\, \oi/\Ha$ and the galaxy masses \Mstar.  The left panels represent all the MGSz objects complying with the above restrictions on S/N, while the right panels represent only the OPARGs from this sample. 

Clearly, those high excitation OPARGs have also the largest \Ha equivalent widths and luminosities of all radio galaxies, and even of Seyfert galaxies in general, and the largest \oiii/\oii ratios. They also belong to the class of objects with large  \Heii/\Hb ratios. These properties argue for a harder ionizing radiation field and a larger ionization parameter (ratio of radiation pressure and gas pressure). These objects do not have prominent \oi lines, which excludes the existence of of a large warm neutral zone produced by X-ray photons. The bottom panel shows that these objects do not have anything special regarding their stellar masses. Nor did we find any specificity of these objects regarding their radio morphologies or black hole masses.

\begin{figure}  
  \centering
    \includegraphics[width=1\linewidth, trim=10 30 20 0]{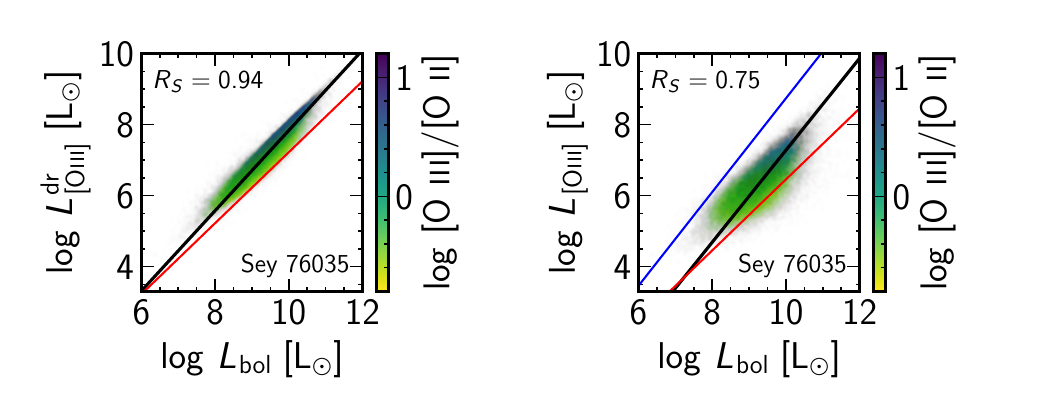}
  \caption{Simple proxies for \Lbol for Seyfert galaxies. Left: \Loiii corrected for extinction; right: \Loiii without extinction correction. The thick black line represents our regressions as given by Eq.~\ref{eq:lbollo3dr} (left panel) and Eq.~\ref{eq:lbollo3} (right panel) considering all 77\,071 Seyferts with \Lbol and \Loiii values (some which are not drawn on the panels above due to the \oii selection cut). In the left panel the red line represents the relation of \citet{Kauffmann.Heckman.2009a}, while in the right panel it represents the relation from \cite{Heckman.etal.2004a}. The blue line in the right panel represents the relation of \citet{Spinoglio.FernandezOntiveros.Malkan.2024a} for Seyfert 2 galaxies.
  }
  \label{fig:lbolproxseyferts} 
\end{figure}

\begin{figure} 
  \centering
  \includegraphics[width=1\linewidth, trim=10 30 20 0]{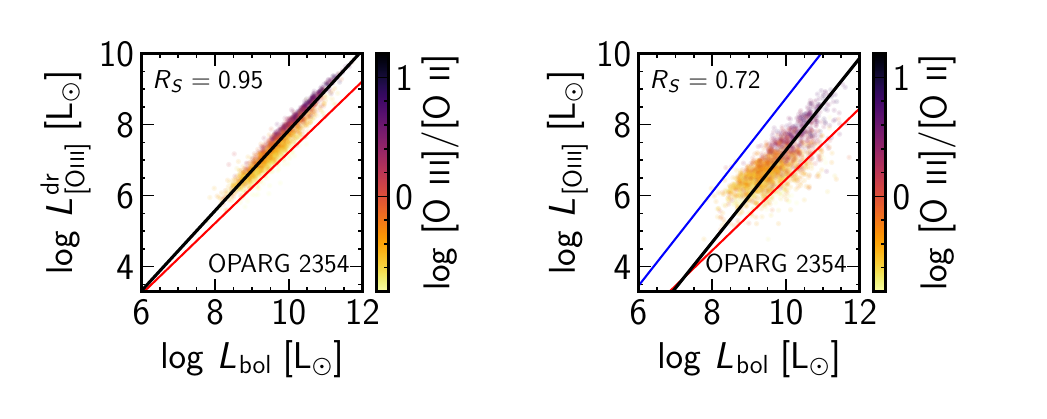}
  \caption{Simple proxies for \Lbol for OPARGs. Left: \Loiii corrected for extinction; right: \Loiii without extinction correction. The black, red and blue lines have the same meaning as in Fig. \ref{fig:lbolproxseyferts} }
  \label{fig:lbolproxRG}
\end{figure}

\section{The Eddington ratios of OPARGs and of radio quiet AGNs}
\label{Eddington}


\subsection{Bolometric luminosities}
\label{boloparg} 

One of the basic parameters describing AGNs are their bolometric luminosities, \Lbol, that is to say the total power of the radiation emitted by the active nucleus. In Type II AGNs, such as considered in this paper, the radiation coming directly from the AGN is obscured by the dusty torus, and one has to use proxies. \citet{Heckman.Best.2014a} propose a list of possible proxies. The only ones accessible for all the AGNs of our sample are those based on optical data. The most popular is the luminosity in the \oiii line, \Loiii. \citet{Kauffmann.Heckman.2009a} and \citet{Heckman.Best.2014a} adopted a bolometric correction of 600 for the extinction-corrected \Loiii luminosity for all their objects. \citet{Netzer.2009a}, \citet{KozielWierzbowska.Stasinska.2011a} and \citet{Sikora.etal.2013a} argued for a use of \LHa rather than \Loiii, since the intensity of \oiii depends on the ionization state of the gas. 

In Appendix \ref{bolcor} we estimate the bolometric correction using AGN photoionization models and taking into account the contribution of \hii regions to the observed emission-line intensities. The models have as an input the spectral energy distributions used by \citet{Ferland.etal.2020a} which are based on the observed continua of unobscured Type I AGN in the optical and X-rays \citep{Jin.Ward.Done.2012a, Jin.etal.2018b}. 

The photoionization models assume a covering factor of 1. Estimates of the covering factor can be obtained from high spectral resolution maps \citep{Capetti.VerdoesKleijn.Chiaberge.2005a, Baldi.Capetti.Giovannini.2019a} or by comparing the thermal infrared emission to the primary AGN radiation \citep{Maiolino.etal.2007a, Toba.etal.2021a}. While previous studies suggested that in some AGNs the covering factor depends on the AGN luminosity \citep[\eg][]{Maiolino.etal.2007a, Netzer.etal.2016a}, \citet{Stalevski.etal.2016a} find that the covering factor depends more weakly on the luminosity than previously thought, with values in range 0.6--0.7. In the following we adopt a covering factor of 0.65.
So we adopt \Lbol = \Lbolmod / 0.65, where \Lbolmod is the bolometric luminosity of the AGN derived in Appendix \ref{bolcor}.

It turns out, which was not obvious from start,  that there is a very strong correlation between the extinction-corrected \Loiii and \Lbol as shown by the left panel of Fig.~\ref{fig:lbolproxseyferts} so \Loiii can indeed be used as a proxy for \Lbol. 
The regression line is plotted with a thick black line. Its equation is:
\begin{equation}
\label{eq:lbollo3dr}
\log \Lbol = (0.8844 \pm 0.0011) \log \Loiii^\mathrm{dr} + (3.0703 \pm 0.0076).
\end{equation}
The figure also shows the relation  from \cite{Kauffmann.Heckman.2009a}:  $\Lbol = 600 \Loiii^\mathrm{dr}$. It can be seen that their expression gives much larger values of the bolometric luminosity than our procedure.

In case where the emission lines cannot be corrected for extinction, one can also use the observed values of \Loiii as a proxy for \Lbol. In this case the uncertainty is larger, but is still quite acceptable. The results are shown in the right panel of  Fig.~\ref{fig:lbolproxseyferts}. 
The regression line, plotted with a thick black line, is now:
\begin{equation}
\label{eq:lbollo3}
\log \Lbol = (0.7715 \pm 0.0023) \log \Loiii + (4.3839 \pm 0.0145).
\end{equation}

The relation of Heckman et al (2004) linking the bolometric luminosity to the uncorrected luminosity in \oiii,  $\Lbol = 3500 \Loiii$, is plotted  in red, while the one derived by \cite{Spinoglio.FernandezOntiveros.Malkan.2024a} is plotted in blue. Our estimation falls between those two estimates!

If interested only in radio galaxies, the same expressions can be used as for AGN galaxies in general, as shown in Fig.~\ref{fig:lbolproxRG}. 

Of course, when both \oiii\Hb and \nii/\Ha are available, it is better to compute \Lbol using the expression  \Lbol = \Lbolmod / 0.65, with \Lbolmod as derived in the Appendix.

The main uncertainties in our determination of the bolometric luminosities lie in the covering factor of the ionizing source by the emission-line gas, the dust-to gas ratio and in the reliability of the adopted SED at high energies.

\subsection{Eddington ratios}
\label{edd}

\begin{figure} 
  \centering
  \includegraphics[width=1\linewidth, trim=20 30 10 0]{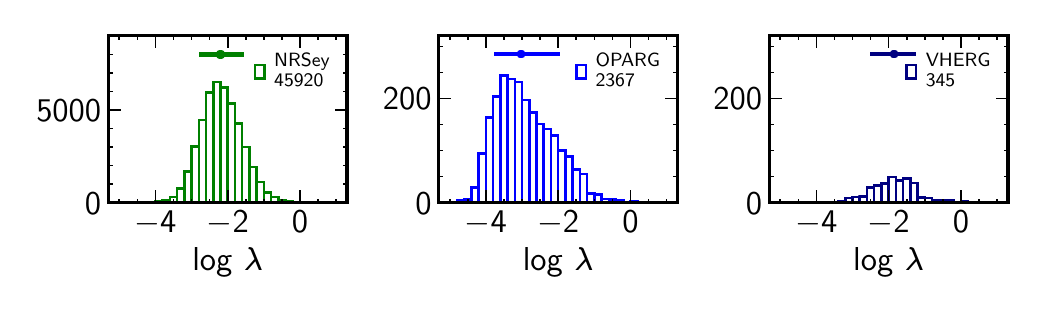}
  \caption{Eddington ratio histograms for non-radio Seyferts (left), 
  OPARGs (middle), and VHERG (right). The horizontal segments on top of the histograms indicate the values of the $(16, 50, 84)$ percentiles. These are $(-2.73, -2.20, -1.61)$ for non-radio Seyferts, $(-3.71, -3.02, -2.02)$ for the OPARGs, and $(-2.46, -1.84, -1.30)$ for VHERGs. }
  \label{fig:edd-histo}
\end{figure}

\begin{figure} 
  \centering
  \includegraphics[width=1\linewidth, trim=15 30 10 0]{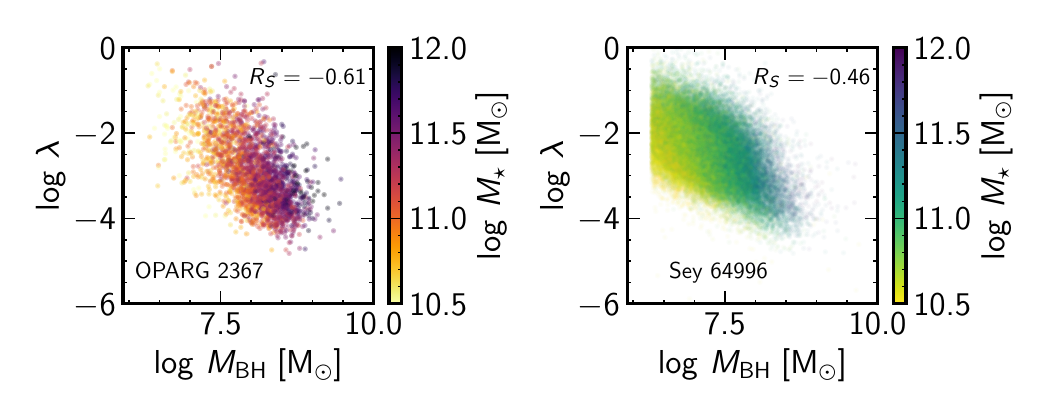}
  \caption{Eddington ratio versus black hole mass, colour-coded by the values of \Mstar. Left panel: OPARGs; right panel: Seyferts.}
  \label{fig:EddrMBH}
\end{figure}

An important parameter quantifying the level of activity of an AGN is its Eddington ratio, $\lambda$, defined as the total radiative power divided by the Eddington luminosity, $L_\mathrm{Edd}$, of the accreting black hole. In practice, we use the expressions $\lambda = \Lbol / L_\mathrm{Edd}$ and $L_\mathrm{Edd} = 1.3 \times 10^{31} \MBH$,  
where \Lbol is in watts and \MBH in solar masses.

Figure~\ref{fig:edd-histo} shows histograms of the Eddington ratios for non-radio Seyferts (left), OPARGs (middle) and VHERGs (see definition below).

We see that the Eddington ratios of the OPARGs are displayed over a very wide range, going as far down as $10^{-5}$.  The $(16, 50, 84)$ percentiles are 
$(-3.71, -3.02, -2.02)$.
Restricting the sample of radio galaxies to objects with \woiii larger than 5~\AA\ (\ie those corresponding to the canonical definition of HERGs), these percentiles  are $(-2.96, -2.19, -1.51)$.
This is at odds with the common view that HERGs have Eddington ratios larger than 0.01 \citep{Buttiglione.etal.2010a, Padovani.etal.2015a, Heckman.Best.2014a}. We have checked that the fact that our sample is much larger than theirs and includes radio galaxies with radio fluxes down to 1~mJy is not the explanation for this difference. Actually, the difference stems mainly from our different estimate of the bolometric luminosities.
Seyferts also exhibit a wide range of Eddington ratios. Note that our Seyferts do include `true' LINERs, but not `fake' AGNs,  that is,  galaxies ionized by their old stellar populations \citep{CidFernandes.etal.2010a}. 

Almost all radio galaxies with Eddington ratios $\lambda$ larger than 0.01   are at the top right  of the BPT diagram as can be seen in Fig.~\ref{fig:BPT-Eddr}. They correspond to the objects with the highest \oiii/\oii ratios, prominent \Heii lines, and the highest values of \LHa and of \wha that were discussed in Sect. \ref{BPT}. Maybe only those objects should be called `high excitation radio galaxies' or, better, `very high excitation radio galaxies': VHERGs? In the following we call VHERGs objects for which $\log\, \oiii/\Hb \ge 0.8$.
Note that these objects constitute the upper branch of the \oiii/\oii versus \wha diagram in Fig.~\ref{fig:WHaWO3O3O2}

A recent study by \citet{Aggarwal.2024a} based on the \citet{Kozlowski.2017a}  catalogue analysed 132,000 objects with redshifts up to 2.5. He finds that at any redshift above 0.7--0.8 there is a clear dependence between the Eddington ratio and the black hole mass. In Fig.~\ref{fig:EddrMBH} we show the values of $\lambda$ as a function of \MBH for our sample of OPARGs (left) and our sample of Seyfert galaxies (right). However, note that the anti-correlation that we see might be driven by the fact that  \MBH appears in both ordinates.

\section{The Eddington-scaled accretion rates of OPARGs and OPIRGs}
\label{Eddaccretion}


\begin{figure} 
  \centering
  \includegraphics[width=1\linewidth, trim=15 30 10 0]{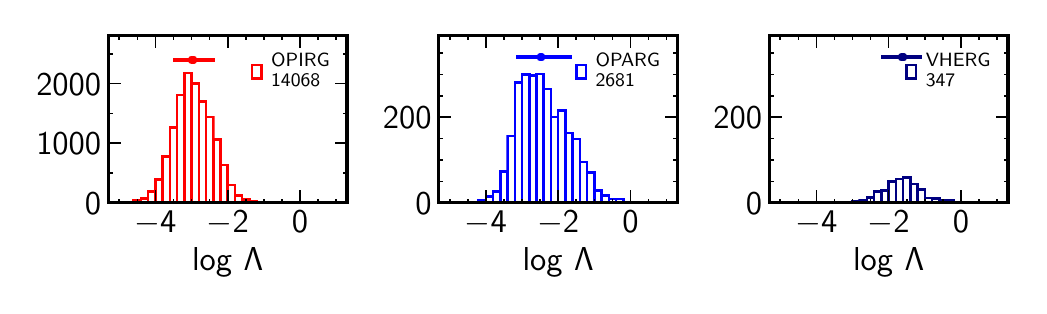}
  \caption{Eddington-scaled accretion-rate histograms for OPIRGs (left), OPARGs (middle), and VHERGs (right). The horizontal segments on top of the histograms indicate the values of the $(16, 50, 84)$ percentiles. These are $(-3.47, -2.97, -2.41)$ for the OPIRGs, $(-3.10, -2.48, -1.67)$ for the OPARGs, and $(-2.16, -1.61, -1.13)$ for the VHERGs. }
  \label{fig:Lambda-histo}
\end{figure}

\begin{figure} 
  \centering
  \includegraphics[width=1\linewidth, trim=20 20 20 0]{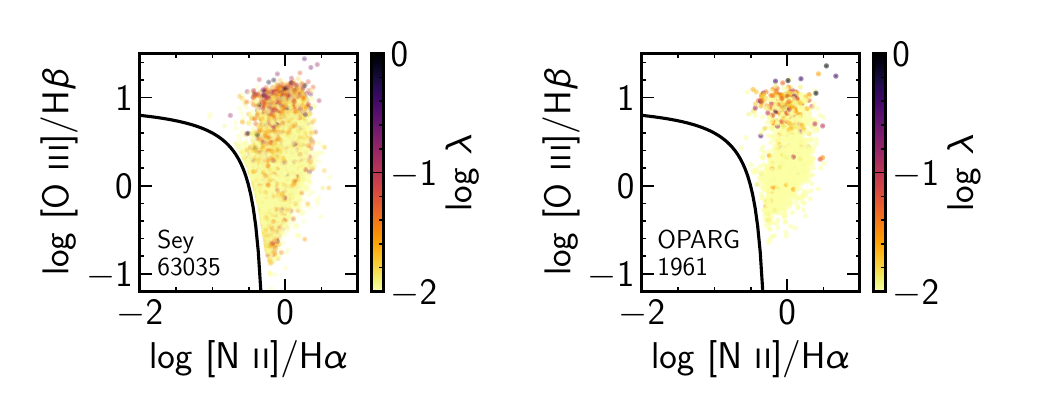}
  \caption{BPT diagram for the AGNs according to the \citet{Stasinska.etal.2006a} line from the MGSz$^\prime$ (left), and for OPARGs (right). Colour-coding is done with respect to the Eddington ratio.}
  \label{fig:BPT-Eddr}
\end{figure}

\begin{figure} 
  \centering
    \includegraphics[width=0.5\linewidth, trim=20 20 20 0]{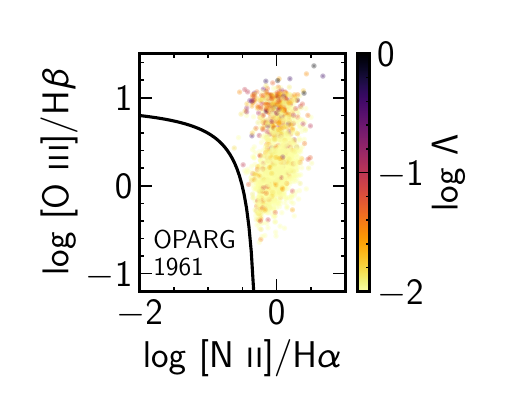}
  \caption{BPT diagram for OPARGs. Colour-coding is done with respect to the Eddington-scaled accretion rate. }
  \label{fig:BPT-Lambda}
\end{figure}
 
Another parameter of importance is the Eddington-scaled accretion rate \citep{Jester.2005a, Best.Heckman.2012a, Williams.etal.2018g, Hardcastle.Croston.2020a, Baldi.etal.2021a, Whittam.etal.2022a}, which is the ratio of the total energetic output of the black hole and the Eddington luminosity. The total energetic output is the sum of the radiative luminosity and of the jet mechanical luminosity.
For the jet mechanical luminosity we adopt the same expression as used by \citet{Best.Heckman.2012a} and introduced by \citet{Cavagnolo.etal.2010a}: $\Lmech $= 7.3 $\times$ $10^{36}$  $(\Lrad/10^{24})^{0.70}$, where \Lrad is in W Hz$^{-1}$ and \Lmech in watts. 
The Eddington-scaled accretion rate $\Lambda$ is thus given\footnote{For OPIRGs we assume that the bolometric luminosity of the AGN is zero, even if \wha is not null.} by $\Lambda = (\Lbol + \Lmech) / \Ledd$.

Figure \ref{fig:BPT-Lambda} shows that our VHERGs are among the objects with the largest values of the Eddington-scaled accretion rates.
 
\citet{Best.Heckman.2012a} find that HERGs typically have Eddington-scaled accretion rates between 0.01 and 0.1, whereas LERGs predominantly accrete at a lower rate. Our results for OPARGs differ from theirs for HERGs because, as can be seen in Fig.  \ref{fig:lbolproxRG}, our estimates of bolometric luminosities give lower values than theirs, and also because our definition of OPARGs includes objects with lower emission-line equivalent widths than their HERGs. Figure~\ref{fig:Lambda-histo} shows the distributions of $\Lambda$  for  the OPIRGs (red), for the OPARGs (blue), and for the VHERGs (navy blue). As in the case of Eddington ratios, the histograms for $\Lambda$ overlap significantly. Most OPIRGs have  $\Lambda$ values smaller than 0.01; however, a lot of OPARGs, and even of the classical HERGs, also have $\Lambda$ values smaller than 0.01.

\begin{figure}
   \centering
   \includegraphics[width=\linewidth, trim=10 25 10 0]{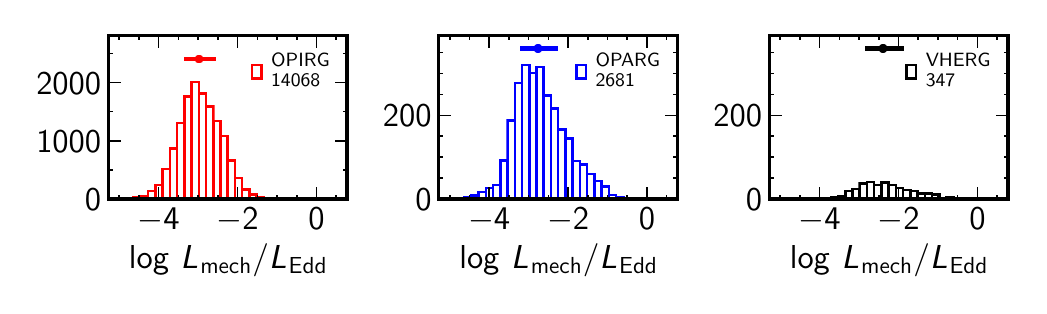}
   \caption{Eddington-scaled mechanical power, $L_\mathrm{mech}/L_\mathrm{Edd}$, histograms for OPIGRs (left), OPARGs (middle), and VHERGs (right).
    The horizontal segments on top of the histograms indicate the values of the $(16, 50, 84)$ percentiles. These are $(-3.47, -2.97, -2.41)$ for the OPIRGs, $(-3.31, -2.76, -2.03)$ for the OPARGs, and $(-2.97, -2.39, -1.63)$ for the VHERGs. 
 }
   \label{fig:Lmech-Ledd}
\end{figure}

In the following, rather than using the Eddington-scaled accretion rate we will use the the Eddington-scaled mechanical rate mechanical power  $L_\mathrm{mech}/L_\mathrm{Edd}$. Figure \ref{fig:Lmech-Ledd} presents the histograms of their distributions for OPIRGs, OPARGs and VHERGs.

\section{Theoretical considerations}
\label{theory}

\begin{table*}[h!]
\centering
\caption{$(16, 50, 84)$ percentiles of the distributions of Eddington-scaled radiative, total and mechanical output, and mechanical-to-radiative luminosity ratio for our subsamples.  } 
\begin{tabular}{lcccc}
\hline \hline
             & $\lambda$               & $\Lambda$               & $L_\mathrm{mech}/L_\mathrm{Edd}$  &  $L_\mathrm{mech}/L_\mathrm{bol}$       \\
\hline
NR-Seyferts  & $(-2.73, -2.20, -1.61)$ & --                      & --                                & --                                      \\ 
OPIRGs       & --                      & $(-3.47, -2.97, -2.41)$ & $(-3.47, -2.97, -2.41)$           & --                                      \\ 
OPARGs       & $(-3.71, -3.02, -2.02)$ & $(-3.10, -2.48, -1.67)$ & $(-3.31, -2.76, -2.03)$           & $(-0.99, 0.13, 0.86)$                   \\ 
VHERGs       & $(-2.46, -1.84, -1.30)$ & $(-2.16, -1.61, -1.13)$ & $(-2.97, -2.39, -1.63)$           & $(-1.08, -0.71, 0.42)$                  \\ 
\hline
\end{tabular}
\label{tab:lambdas}
\end{table*}

In this section we use the results on Eddington ratios and Eddington-scaled mechanical energies obtained for Seyferts, OPIRGs, OPARGs and VHERGs  (see Table~\ref{tab:lambdas}) to pin down the accretion properties and jet launching mechanisms in these objects.

\subsection{Generalities on accretion flows}
\label{9generalities}

When one attempts to describe the accretion regimes in AGNs one should keep in mind that the observed output, either radiative or mechanical, is actually a fraction of the energy of the underlying accretion flow. 
The bolometric luminosity \Lbol associated with the disk radiative output and the mechanical luminosity \Lmech  associated with the formation of jets are thus given by the formula:
\begin{equation}
    L_{\rm output}=c^2\, \eta_{\rm output}\, \dot{M},
\label{Eq:Loutput}    
\end{equation}
where `output' stands either for `bol' or for `mech',  $c$ is the speed of light in the vacuum, $\dot{M}$ denotes the rate at which matter is inflowing into the disk, and $\eta$ should be understood as the efficiency of converting the gravitational energy of accretion into the energetic of the specific output. As we will show in following subsections these efficiencies are highly model-dependent.

We will adopt the standard convention that the radiative efficiency of the disk at the Eddington limit is $\eta_{\rm Edd}=0.1$. 

Thus the mass accretion rate scaled by the Eddington accretion rate can be written:
\begin{equation}
  \dot{m}\equiv\frac{\dot{M}}{\dot{M}_{\rm Edd}}=\frac{0.1}{\eta_\mathrm{output}}\frac{L_{\mathrm{output}}}{L_\mathrm{Edd}}.
\label{Eq:mdot}  
\end{equation}

\subsubsection{High accretion rates ($\mdot >10^{-2}$)}
\label{9highaccr}

In this case, also called cold accretion, accretion occurs through the standard geometrically thin and optically thick disk model, the Shakura-Sunyaev disk \citep[SSD; ][]{Shakura.Sunyaev.1973a}. Part of the gravitational energy of the disk is transformed into thermal energy and is radiated from its surface. Since the disk is optically thick  it is in complete thermodynamic equilibrium. It emits a multi-component black-body spectrum, and most of the energy is radiated in the UV and optical bands.  
The average radiative efficiency for a non-rotating black hole is  $\eta_\mathrm{rad} \sim 6\%$ \citep{Jovanovic.2012a}.
If the black hole  has a non-zero spin the radiation efficiency is larger, the maximum efficiency being $\eta_\mathrm{rad} \sim 30\%$ \citep{Thorne.1974a}. 

Some modifications of this model are possible, for instance when the magnetic field becomes dynamically dominant and the thin disk accretes in a magnetically arrested disk (MAD) regime. Then the accumulated poloidal magnetic field disrupts the accretion flow reducing its kinetic energy and breaking the flow into magnetically confined blobs or streams \citep{Igumenshchev.Narayan.Abramowicz.2003a}.  As a result the radiative efficiency of the thin disk can be modified, but still remaining similar to the values for the non-magnetized thin disk model \citep{Narayan.Igumenshchev.Abramowicz.2003a}. It does not reach the maximum values obtained by \citet{Thorne.1974a}.

\subsubsection{Low accretion rates ($\dot{m} < 10^{-2}$)}
\label{9lowaccr}

In this case, also called hot accretion, the density of the disk is too low for efficient radiation to occur and advection becomes the dominant cooling mechanism. This case  known as radiatively inefficient accretion flow (RIAF)  was first introduced in \citet{Shapiro.Lightman.Eardley.1976a} who proposed a two-temperature accretion disk model to explain the X-ray spectrum of  Cyg X-1. The idea was later expanded in \citet{Narayan.Yi.1994a, Narayan.Yi.1995a} and is known as advection dominated accretion flow (ADAF). 

The radiation processes in ADAF are synchrotron emission and bremsstrahlung, modified by Comptonization \citep{Yuan.Narayan.2014a}. 

The ADAF solution in its classical form (when all the  matter is eventually accreted onto a black hole) is unstable and simulations show that outflows are inevitable.
\citet{Blandford.Begelman.1999a} proposed a  power-law scaling for the mass accretion rate $\dot{M}$, that takes mass loss into account:
 \begin{equation}
 \label{eq:Mdot-R}
\dot{M}(R)=\dot{M}_{\rm BH}\left(\frac{R}{R_{\rm S}}\right)^s,
 \end{equation}
where $R_{\rm S}$ is the Schwarzschild radius, $\dot{M}_{\rm BH}$ is the mass accretion rate at  $R_{\rm S}$, and $s$ is the `mass-loss' index,
Observations indicate that $s$ can vary from $\sim 0.5$ to $\sim 1$ as estimated with various methods \citep{Baganoff.etal.2003a, Marrone.etal.2007a, Wang.etal.2013f, Wong.etal.2011a, Kuo.etal.2014a}.

At moderate accretion rates  ($\mdot \sim 0.01$) the efficiency remains the same as in a standard thin disk ($\eta=0.1$). However, when $\mdot \ll 0.01$ the radiative efficiency of the accretion flows drops quickly and is roughly given by (see \eg \citealp{Narayan.McClintock.2008a}): 
 \begin{equation}
\eta\sim0.1 \frac{\dot{M}}{0.01\dot{M}_\mathrm{Edd}} .
\label{eq:etaADAF}
 \end{equation}
Keeping in mind our considerations from the beginning of this section  one can obtain the relation between $\dot{m}$ and $\lambda$ for $\lambda<0.01$:
 \begin{equation}
   \dot{m}\sim0.1\sqrt{\lambda}.
   \label{eq:mlambda}
 \end{equation}

There are actually three  regimes of ADAF (see  \citealp{Yuan.Narayan.2014a}). 

For the highest accretion rates ($\dot{m} \sim 10^{-2}$--$10^{-3}$),  
radiation losses dominate  and  the radiative efficiency decreases rapidly with decreasing accretion rate. This solution is known as luminous hot accretion flow (LHAF). The radiative efficiencies become comparable to those estimated for a geometrically thin disk, although the gas remains hot due to compressional heating.

At lower accretion rates  the electrons can efficiently radiate their viscous energy as well as the small amount of energy gained through collisions with ions. However, because the Coulomb collisions are inefficient, the system is advection dominated. The radiative efficiency is still high and can be of the order of one to a few percent. 

At the lowest accretion rate ($\dot{m}<10^{-4}$) even electrons cannot efficiently radiate and are advection-dominated. The system is truly radiatively inefficient and this regime is called electron ADAF (eADAF). It is characteristic of the dimmest accretion systems.

\subsubsection{Intermediate accretion rates}
\label{9interaccr}

The transition from radiatively efficient to inefficient flow is not abrupt and both regimes can coexist in the same disk. The region in which the ADAF solution dominates is a function of the radius and mass accretion rate. This transition however is hardly probed in general relativistic magnetohydrodynamical (GRMHD) simulations. It is expected that the ADAF model at moderately high accretion rates ($\dot{m} < 0.01$) will form in the inner region of the accretion disk, while the geometrically thin and optically thick solution is still possible in the outer zone of the disk \citep{Esin.McClintock.Narayan.1997a, Narayan.McClintock.Yi.1996a, Narayan.Mahadevan.Quataert.1998a, Yuan.Narayan.2014a}. Obviously the extent of the ADAF solution increases with decreasing accretion rates.

The weak gravitational bonding of the gas in the thick accretion flow can easily produce outflows. Several variants of the ADAF solution are found.  One  of them is the adiabatic inflow-outflow solution (ADIOS).
The presence of outflow/winds will only further decrease the radiative efficiency of the disk.

\subsection{Generalities on jets}
\label{9generaljets}

Just as we discussed in Sect.~\ref{9generalities} considering the bolometric luminosity, the mechanical jet luminosity is proportional to its production efficiency and the mass accretion rate (Eq.~\ref{Eq:Loutput}). However, unlike the case of radiative efficiency which is strictly related to the disk internal physics, the jet efficiency can span several orders of magnitude depending on the way the jet energy is extracted and it can even exceed the available accretion energy.

Jets are considered to be powered by two types of processes. \citet{Blandford.Znajek.1977a} proposed a process  where the energy is extracted from a spinning black hole in the presence of a large-scale poloidal magnetic field attached to the black hole horizon. In the \citet{Blandford.Payne.1982a} process, the jet is strictly powered by the differential rotation between the accretion disk and the poloidal magnetic field attached to the disk.

\subsubsection{Jet production of thin disks}
\label{9jetthin}

The jet production efficiency in the standard thin disk model is expected to be low but can be elevated if the disk is subject to magnetically driven winds/jets \citep{Li.Cao.2012a, Li.Begelman.2014a}. If a sufficiently strong poloidal magnetic field is accumulated in the inner region of the disk then for a maximally spinning black hole the jet production efficiency can reach $\eta_\mathrm{mech} \sim 10\%$. However this would result in a decrease of the disk luminosity by more than two orders of magnitude \citep{Li.2014a} since part of the energy is lost in outflows. If the magnetic field is moderately high the disk radiative efficiency will be unaffected while $\eta_\mathrm{mech}$ would become of the order of $3$--$10\%$.

\subsubsection{Jet production of thick disks}
\label{9jetthick}

\citet{Nemmen.Tchekhovskoy.2015a} showed that, in a sample of nearby low luminosity AGNs, the median jet efficiency  $\eta_\mathrm{mech}$ based on the ADAF model is of the order of a few percent. However, they note that  accretion rates estimated under the ADAF model should be considered as upper limits since this model does not take outflows into account. Consequently the kinetic efficiencies computed under ADAF are lower limits. 

Using Eq.~\ref{eq:Mdot-R} they then computed the median kinetic efficiency with the ADIOS model as a function of the mass loss index $s$ and find 
\begin{equation}
\overline{\eta_\mathrm{mech}} = 2.8 \cdot 10^{4s}\,\, \%.
\end{equation}
When computed in the framework of the ADIOS model the jet production efficiencies in these low luminosity AGNs  are in the range  $100$--$1000\%$. This by far exceeds the maximum values obtained in GRMHD simulations for maximally spinning black holes accreting in the MAD regime \citep{Tchekhovskoy.Narayan.McKinney.2011a, McKinney.Tchekhovskoy.Blandford.2012a}.

\subsection{The nature of our different classes of galaxies containing AGNs}
\label{9nature}

The theoretical considerations in the preceding subsections were based on the important parameters $\dot{m}$ and  $L_{\rm mech}$. These quantities are however not available directly from the observations, so to infer the accretion and jet production mechanisms in the objects studied in this paper, we use the values of $\lambda$ and Eddington-scaled mechanical energies as proxies. 

\subsubsection{OPIRGs}
\label{9opirgs}

The OPIRGs in our sample are thought to accrete at a radiatively inefficient regime. If we adopt the ADAF model,  the observed range of values of the Eddington-scaled mechanical power $L_{\mathrm{mech}}/{L_\mathrm{Edd}}$ (see Table \ref{tab:lambdas}) with an average jet efficiency of $3\%$ imply that  $\dot{m}$   lies in the range $10^{-3}$--$10^{-2}$. 
 
Such range of values is unlikely, as high radiative efficiency of the ADAF model in this regime would produce the spectrum able to excite the gas in the vicinity of the AGN and emission lines would be seen. This is however not the case in OPIRGs which do not show any optical AGN signatures.
If the accretion is described by the ADIOS model, jets have typical efficiencies $\eta_\mathrm{mech} >100\%$ (going as high as $\eta_\mathrm{mech}\sim 1000\%$). If we assume such values for our sample of OPIRGs and given the range of values of $L_\mathrm{mech}/L_\mathrm{Edd}$ shown in Table \ref{tab:lambdas}, the values of $\dot{m}$ range from $\sim10^{-5}$ to $\sim10^{-4}$ .

Thus the OPIRGs emission is fully consistent with a radiatively inefficient flow. Based on our previous considerations we can claim that the emission of OPIRGs is fully consistent with the model of truly radiatively inefficient radiation flow (\ie electron ADAF).

\begin{figure}
    \centering
    \includegraphics[width=\linewidth, trim=10 30 20 0]{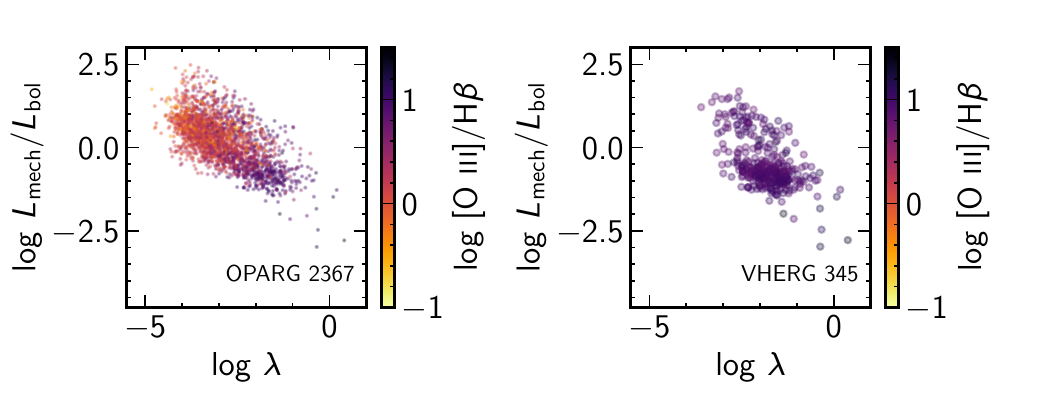}
    \caption{Mechanical to bolometric luminosity ratio as a function of the Eddington ratio colour-coded by $\oiii/\Hb$ for OPARGs (left) and VHERGs (right).
    }
    \label{fig:efficiencies}
\end{figure}

\subsubsection{OPARGs}
\label{9opargs}

From the observed range of $\lambda$ in Table 1 we can calculate the mass accretion rate using the simplified formula derived for ADAF and given by Eq.~\ref{eq:mlambda}. We find that the $16$--$84$ percentiles for $\dot{m}$ would be $10^{-2.85}$--$10^{-2.01}$. According to our considerations in Sect. \ref{9lowaccr} such values correspond to the ADAF model in the LHAF regime for $\dot{m} > 10^{-2}$ and to the classical ADAF regime (\ie intermediate between LHAF and eADAF) for the majority of OPARGs which have lower values of $\dot{m}$. We note that at such accretion rates the outer thin disk still can exist and the expected radiative efficiency will be increased. The resulting mass accretion rate can be slightly lower than the values reported here.
The corresponding radiative efficiencies estimated from Eq.~\ref{eq:etaADAF} in this mass accretion rate range are in the $1$--$10\%$ range.

In the left panel of Fig.~\ref{fig:efficiencies} we plot the values of the mechanical to bolometric luminosity ratio as a function of the Eddington ratio colour-coded by $\oiii/\Hb$. This corresponds to the ratio of the mechanical to the radiative efficiency (see Eq.~\ref{Eq:Loutput}). Objects of lower excitation have a larger mechanical to radiative output ratio. We note that the mechanical luminosities should be understood as an average over the total jet launching phase. On the other hand the bolometric luminosities, estimated from optical lines, correspond to the present phase of accretion. Thus a large scatter in Fig.~\ref{fig:efficiencies} is observed. In general, if the Eddington ratio increases, the mechanical output decreases with respect to the radiative output but some exceptions are possible (\eg if the Eddington ratio was elevated in a recent accretion event).

 We see that OPARGs at $\lambda<10^{-2}$ have mechanical efficiencies comparable to their radiative efficiencies or higher, up to an order of magnitude larger for smallest value of $\lambda$, where $\eta_{rad}$ is of the order of one to a few percent. This considerations yields mechanical efficiencies of up to 10\%.
 
\subsubsection{VHERGs}
\label{9vhergs}

From the observed range of $\lambda$ in Table~\ref{tab:lambdas} for the subpopulation of OPARGs that we called VHERGs and assuming $\eta = 10\% $,   Eq.~\ref{Eq:mdot}  leads to a range for  $\dot{m}$ of $10^{-2.46}$--$10^{-1.30}$ and a median value of $10^{-1.84}$. This is compatible with thin disk accretion for the majority of VHERGs. For those with $\lambda < 10^{-2}$, the model would be ADAF in the central parts and SSD in the outer parts of the accretion disk.
The right panel of Fig.~\ref{fig:efficiencies} shows that most of the VHERGs  produce very inefficient outflows with mechanical efficiencies at most comparable to their disk radiative efficiencies.

\subsubsection{Non-radio Seyferts }
\label{9optagn}

As seen in Table~\ref{tab:lambdas} most Seyferts that are not radio-AGN have Eddington-scaled accretion rates in the range $10^{-2.74}$--$10^{-1.61}$. Thus from Sect. \ref{9highaccr} the accretion mode is that of a thin disk, except possibly an ADAF mode in the central parts for the objects with the lowest Eddington ratios. We note that the distribution of  Eddington ratios of non-radio Seyferts peaks at a higher value ($\log \lambda = -2.20$) than that of OPARGs ($\log \lambda = -3.02$). This could be due to a selection effect. Seyfert galaxies having  high accretion rates are expected to have very low jet production efficiency  and thus  may stay undetected in flux-limited radio surveys. In addition, being  mostly spiral galaxies, Seyferts generally lack the poloidal component of the magnetic field to form outflows efficiently so the accretion energy is mainly released by radiation, which is a further reason for them  to remain undetected in radio.

\begin{figure} 
 \centering
 \includegraphics[width=1\linewidth, trim=10 25 10 0]{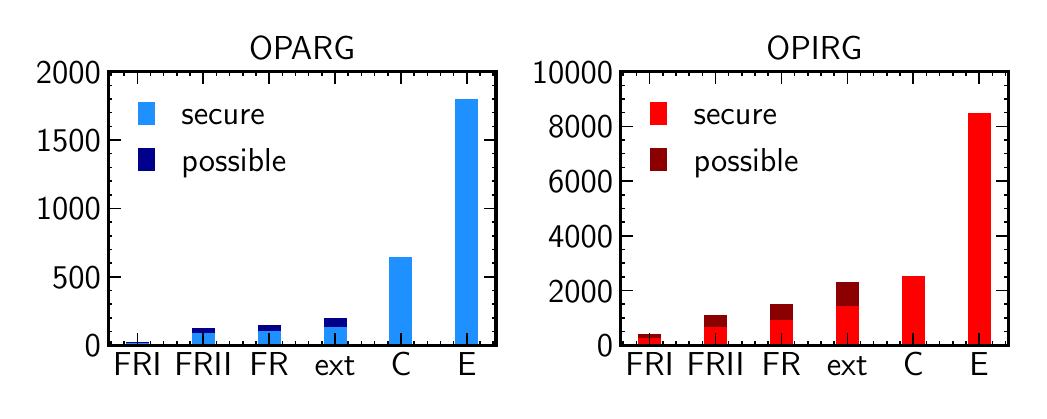}
 \caption{Classification of OPARGs (left) and OPIRGs (left) into radio morphological types according to the ROGUE I and II catalogues. FRI stands for \citet[FR]{Fanaroff.Riley.1974a} type I, FRII for FR type II, FR for FR types I, II and I/II, ext for all extended types (FR, OI, OII, Z, X, DD, WAT, NAT and HT), C for compact, and E for elongated (see \citealp{KozielWierzbowska.Goyal.Zywucka.2020a} for details on the radio morphologies). `Possible' classifications are in a darker colour to contrast with `secure' classifications.  }
 \label{fig:morph}
\end{figure}

\subsubsection{Final comments}
\label{9final} 

\begin{figure*} 
 \centering
 \includegraphics[width=1\linewidth, trim=0 30 0 0]{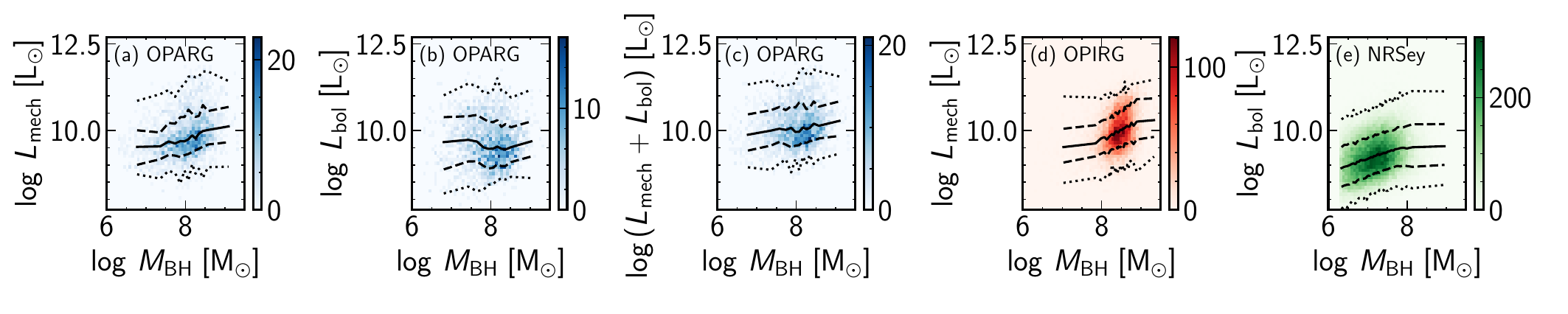}
 \caption{2D histograms of energy outputs versus black hole mass for the different categories of AGN galaxies. The two leftmost panels show (a) the mechanical energy output for OPARG and (b) the radiative energy output for OPARGs. The three rightmost panels show the total energy output for (c) OPARGs, (d) OPIRGs and (e) non-radio Seyferts.  
 The  colour-scale gives the number of objects in each bin. The curves show the value of the median (solid line), the 16 and 84 percentiles (dashed lines) and 1 and 99 percentiles (dotted lines) as a function of black hole mass.  }
  \label{fig:Ltotvsmbh}
\end{figure*}

To sum up our findings, Table \ref{tab:efficiencysummary} shows the estimated mass accretion rates $\dot{m}$, radiative efficiencies of the accretion flow $\eta_\mathrm{rad}$, and the mechanical efficiencies $\eta_\mathrm{mech}$ for OPARGs and OPIRGs. OPARGs$_{100}$ corresponds to mass accretion rates estimated assuming a mechanical efficiency $\eta_{mech}\approx 100\%$, and OPARGs$_{1000}$ corresponds to mass accretion rates estimated assuming a mechanical efficiency $\eta_\mathrm{mech}\approx 1000\%$.
Our interpretation of the nature of the flows in OPIRGs and OPARGs is well in agreement with the observed contrast in the distributions for radio morphologies for OPIRGs and OPARGs. In Fig. \ref {fig:morph} it is found that  16\% of the OPIRGs are associated with extended radio morphologies (which are mostly of FR I and FR II types) while this fraction drops to 7.4\%  for OPARGs.  Moreover we note that, due to the fact that the mechanical power represents an average of jet production efficiencies over time, this mechanism may be not related to current episodes of activity, especially for the most extended radio galaxies. This could explain why we observe powerful radio galaxies in sources with extremely high values of $\lambda$. We postulate that the increase of radiative efficiency in those sources can be due to a recent episode of increased accretion rate.  

One of the the results of our comparison of OPIRGs and OPARGs in Sect.~\ref{populations} was that the distributions of their radio luminosities are almost identical (see Figs.~\ref{fig:Vmax} and \ref{fig:compare-OPARG-OPIRG}). We believe that the explanation lies in the fact that on average OPARGs have mass accretion rates two orders of magnitude higher than OPIRGs but mechanical efficiencies two orders of magnitude lower so that the two effects roughly compensate (see Table~\ref{tab:efficiencysummary}).

\begin{table}[]
    \centering
    \caption{Estimated values of  $\dot{m}$,  $\eta_{rad}$, and  $\eta_{mech}$ for  OPARGs and OPIRGs. }
    \begin{tabular}{cccc}
    \hline \hline
                & $\dot{m}$ &$\eta_\mathrm{rad}$ &$\eta_\mathrm{mech}$\\
    \hline
        OPARGs & $10^{-2.85}$--$10^{-2.01}$ & $1$--$10\%$ & $1$--$10\%$ \\
        OPIRGs$_{100}$ & $10^{-4.47}$--$10^{-3.41}$& $\ll 1\%$ & $100\%$\\
        OPIRGs$_{1000}$ & $10^{-5.47}$--$10^{-4.41}$& $\ll 1\%$ & $1000\%$ \\
    \hline
    \end{tabular}
    \label{tab:efficiencysummary}
\end{table}

As a final glimpse on radio and non-radio AGNs in the SDSS, we show in  Fig.  \ref{fig:Ltotvsmbh} the values of the total energy output of the AGNs as a function of black hole mass in the form of 2D histograms. These plots illustrate the fact that the energy output is only mildly dependent on the black hole mass in radio galaxies, especially in OPARGs. It also shows that, for overlapping black hole masses, radio galaxies have higher total energy output than non-radio Seyferts.

\section{Summary}
\label{sum}

We used the ROGUE I and II catalogues (\citealp{KozielWierzbowska.Goyal.Zywucka.2020a} and 2024 in preparation), the largest human-made catalogues of radio sources associated with optical galaxies, to revisit the characterization of radio AGNs. We focussed on their radio luminosities and on properties derived from the analyses of the SDSS spectra of their associated galaxies.

In order to deal with a complete sample, for this work we selected only the objects pertaining to the SDSS MGS sample, which is virtually complete down to a magnitude $m_r$ of 17.77. We also applied the redshift limits 0.002, so that luminosity distances are reliable, and 0.4, to ensure that the \Ha line lies in the observable wavelength range.

The ROGUE catalogues make no claim about the origin of the radio emission. To extract the radio AGNs, we used the $D_n(4000)$ versus $L_{1.4}/$\Mstar diagram, which has been shown to classify all the sources of our starting sample with a very high reliability \citep{KozielWierzbowska.etal.2021a}. 

Our final sample contains 16803 radio AGNs. This is much larger than the sample of \citet{Best.Heckman.2012a}, which is the basis of many statistical studies on radio galaxies, and which contains only 7302 radio AGNs. This large difference comes mainly from the fact that in the ROGUE catalogues, the cross-matching goes down to a flux density level of 1~mJy, while in \citeauthor{Best.Heckman.2012a} it goes down to 5~mJy only.

In view of the fact that there is no full agreement on the way to separate  radio galaxies into LERGs and HERGs, and  that the criteria are not physically based, we propose another classification, based on the equivalent width of \Ha, \wha. It has been shown by \citet{CidFernandes.etal.2010a, CidFernandes.etal.2011a} that, statistically, SDSS galaxies that have a \wha smaller than 3 \AA\ are not ionized by an AGN but by their own population of HOLMES. We thus call  those radio galaxies that have a $\wha <3$~\AA\  `optically
inactive'. The remaining ones are called `optically active' radio galaxies (OPARGs as opposed to OPIRGs). In our sample, there are 14082 OPIRGs and 2721 OPARGs.

We then compared the global properties of OPIRGs and OPARGs. For this we had to correct for the selection effects due to the fact that the SDSS spectroscopic catalogue is magnitude-limited.

We find that  the distributions of radio luminosities \Lrad of OPARGs and OPIRGs are virtually undistinguishable. The same actually also occurs for HERGS and LERGs in our sample, while \citet{Heckman.Best.2014a} consider HERGs to have higher radio luminosities than LERGs.
On average, the black hole masses of OPIRGs are larger than those of OPARGs and so are the values of the total stellar masses \Mstar.
The spectral discontinuity, $D_n(4000)$, the mean stellar age, and stellar extinction, $A_V$,  all suggest that OPARGs experience some level of star formation. This interpretation is in line with that of previous studies on HERGs and LERGs, even if  HERGs constitute only a small sub-population of OPARGs.  This argues in favour of our OPIRG-OPARG classification, based on physical arguments. 

We find that, as regards to stellar masses, OPARGs can be viewed as the extension of radio galaxies towards higher masses of Seyferts, and OPIRGs as the extension of retired galaxies.
The proportion of radio galaxies in the total MGSz sample is 1\%.

Plotting the OPARGs in the BPT diagram, we compared their distribution with that of the remaining galaxies. We find that there is a sub-family of very high excitation radio galaxies (which we called VHERGS)  at the top of the AGN wing. This group is slightly displaced towards the left of the rest of the AGN galaxies, suggesting a higher ionization parameter, meaning a stronger ionizing radiation field with respect to the gas pressure. 
 
To compare the Eddington ratios of OPARGs with those of Seyferts, we first devised a method to obtain the bolometric luminosities of these objects from the data at hand. This method, which uses the observed \Ha luminosity and the position of the objects in the BPT diagram, is extensively described in Appendix \ref{bolcor}. It takes into account the contribution of young stars to the observed line emission. It turns out that, contrary to previous claims, the resulting bolometric luminosity is very strongly correlated with the luminosity in the \oiii line (the correlation being almost perfect if  the \oiii luminosity has been corrected for extinction). We provide formulae to derive bolometric luminosities from \Loiii. They differ  significantly from formulae proposed by other groups on different grounds. For example, our bolometric luminosities are smaller than those estimated by \citet{Heckman.etal.2004a} and larger than those estimated by \citet{Spinoglio.FernandezOntiveros.Malkan.2024a} for Seyfert 2 galaxies. Because of this discrepancy, we find lower Eddington ratios and Eddington-scaled accretion rates than those found, for example, by \citet{Heckman.Best.2014a} for radio galaxies.

We find that only a small group of VHERGs have Eddington ratios and Eddington-scaled accretion rates higher than $10^{-2}$, which is canonically considered as the lower limit for the occurrence of radiative efficient accretion \citep{Heckman.Best.2014a, Netzer.2015a, Padovani.etal.2017a}. 
If our estimates of the bolometric luminosities are correct, this means than only a small proportion of mainstream HERGs are indeed radiatively efficient.

From theoretical considerations on accretion mechanisms in AGNs, we suggest that in OPARGs, with their moderate to high accretion rates,  the accretion disk is emitting with a thermal spectrum possibly modulated with the spectrum of an advection-dominated accretion flow and this produces observable emission lines. In VHERGs, which constitute the high-excitation extreme of the OPARG population, the accretion occurs with very low mechanical efficiencies, but high radiative efficiencies. OPIRGs, on the other hand, are characterized by extremely low radiative efficiency but very high mechanical efficiency, consistent with the existence of truly radiatively inefficient radiation flows.


\begin{acknowledgements}

This work was done within the framework of the research project no.\ 2021/43/B/ST9/03246 financed by the National Science Centre/Narodowe Centrum Nauki. 
We thank Christophe Morisset for his invaluable help with the photoionization models, and Maria Eduarda Ramos Pedro, Lis Cristine Fortes and Janayna de Souza Mendes for their help in structuring the first version of our model grids.
NVA acknowledges support of Conselho Nacional de Desenvolvimento Cient\'{i}fico e Tecnol\'{o}gico (CNPq). GS and NVA acknowledge financial support from the Jagiellonian University. AW was supported by the GACR grant 21-13491X.

Funding for the Sloan Digital Sky Survey V has been provided by the Alfred P. Sloan Foundation, the Heising-Simons Foundation, the National Science Foundation, and the Participating Institutions. SDSS acknowledges support and resources from the Center for High-Performance Computing at the University of Utah. SDSS telescopes are located at Apache Point Observatory, funded by the Astrophysical Research Consortium and operated by New Mexico State University, and at Las Campanas Observatory, operated by the Carnegie Institution for Science. The SDSS web site is \url{www.sdss.org}.

SDSS is managed by the Astrophysical Research Consortium for the Participating Institutions of the SDSS Collaboration, including Caltech, The Carnegie Institution for Science, Chilean National Time Allocation Committee (CNTAC) ratified researchers, The Flatiron Institute, the Gotham Participation Group, Harvard University, Heidelberg University, The Johns Hopkins University, L’Ecole polytechnique f\'{e}d\'{e}rale de Lausanne (EPFL), Leibniz-Institut f\"{u}r Astrophysik Potsdam (AIP), Max-Planck-Institut f\"{u}r Astronomie (MPIA Heidelberg), Max-Planck-Institut f\"{u}r Extraterrestrische Physik (MPE), Nanjing University, National Astronomical Observatories of China (NAOC), New Mexico State University, The Ohio State University, Pennsylvania State University, Smithsonian Astrophysical Observatory, Space Telescope Science Institute (STScI), the Stellar Astrophysics Participation Group, Universidad Nacional Aut\'{o}noma de M\'{e}xico, University of Arizona, University of Colorado Boulder, University of Illinois at Urbana-Champaign, University of Toronto, University of Utah, University of Virginia, Yale University, and Yunnan University.

This research made use of Astropy,\footnote{Astropy Python package: \url{http://www.astropy.org}} a community-developed core Python package for Astronomy \citep{AstropyCollaboration.etal.2013a, AstropyCollaboration.etal.2018a}.
 
\end{acknowledgements}


\bibliographystyle{aa}
\bibliography{references}

\clearpage
\begin{appendix}

\section{Fraction of low \wha galaxies with an active nucleus}
\label{MANGA}

Among galaxies with $\wha < 3$\,\AA\ there may be objects where the \Ha emission is not due to HOLMES but to a weak AGN. 
In order to quantify this effect, we have selected galaxies from the MaNGA Data Release 15 \citep{Aguado.etal.2019a} in the same way as \citet[see their Sect.~2.3]{ValeAsari.etal.2019a}. MaNGA cubes, with an average resolution of $\sim 1$\,kpc, allow us to see whether there is a peak in \wha in the nucleus of a galaxy, which can be attributed to an weak AGN.

\begin{figure}[b]
  \centering
    \includegraphics[width=.9\columnwidth, trim=0 0 0 0, clip]{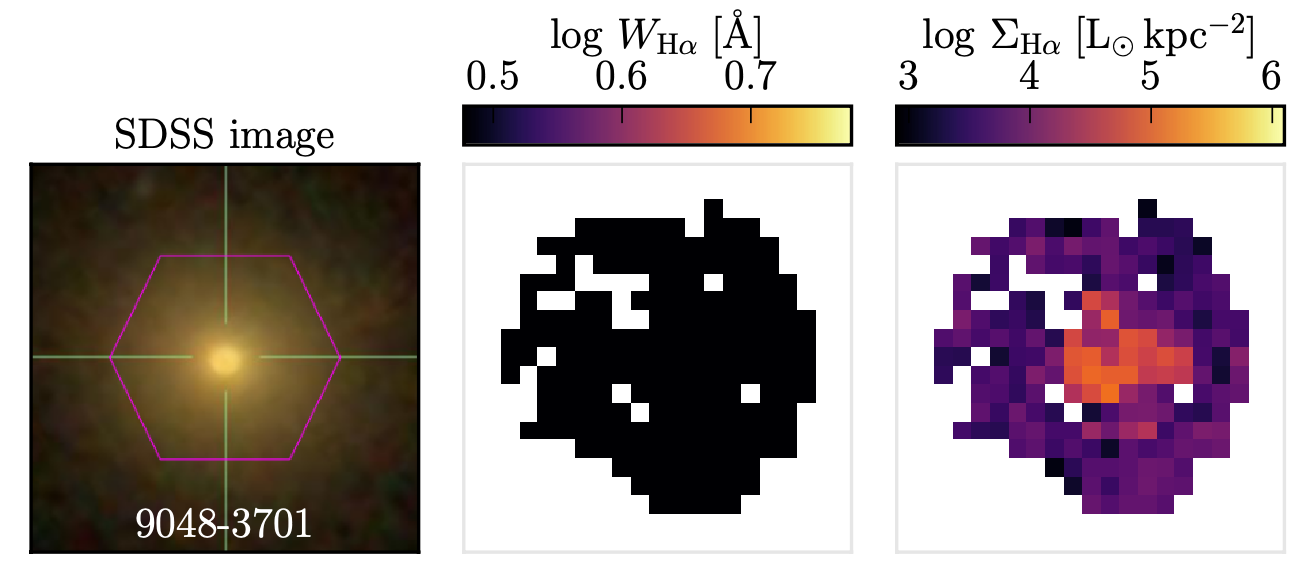}
    \includegraphics[width=.9\columnwidth, trim=0 0 0 0, clip]{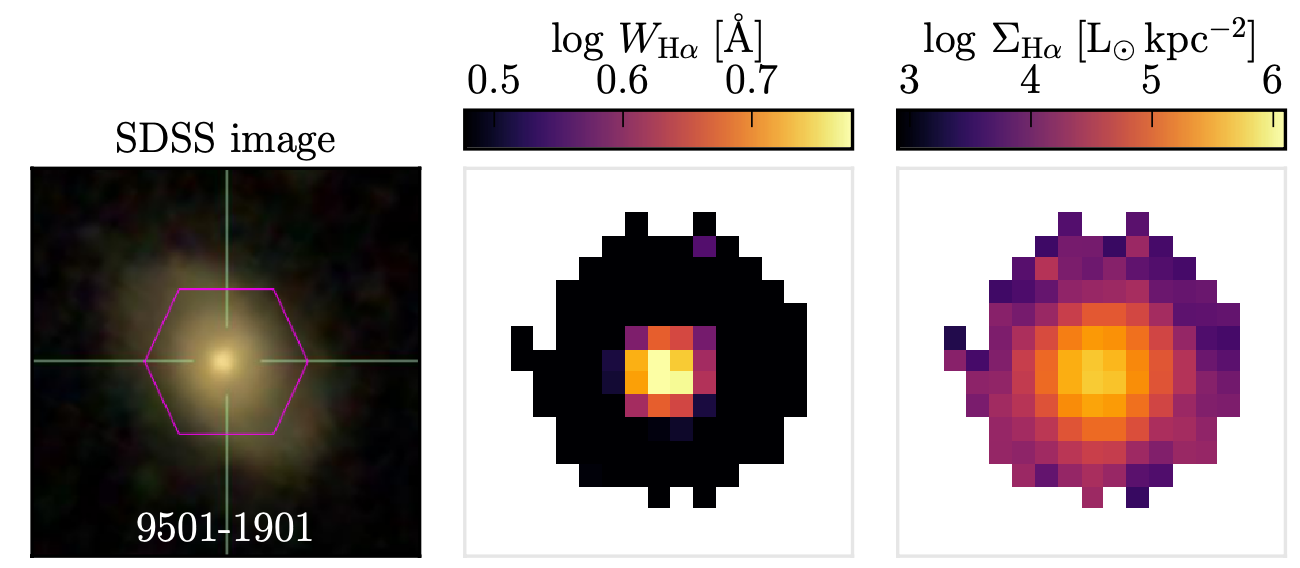}
    \caption{Examples of MaNGA galaxies with integrated $\wha < 3$\,\AA. Panels from the left to the right show the SDSS optical image, the MaNGA \wha map, and the MaNGA \Ha surface density map. The galaxy at the top is an example of an object with globally low \wha, whereas the galaxy at the bottom shows a peak in \wha in its nucleus suggesting a weak AGN.
    }
  \label{fig:manga-agn-holmes}
\end{figure}

From a parent sample of 3\,236 objects, we have selected all 1\,411 objects with integrated $\wha < 3$\,\AA\ (\ie stacking all spaxels in the MaNGA data cube). We have then searched for a peak in \wha near the nuclear region by looking at both their SDSS optical images and \wha maps. We have found evidence of a weak AGN in 55 galaxies, which correspond to 4\% of the $\wha < 3$\,\AA\ sample. Figure~\ref{fig:manga-agn-holmes} shows an example of a galaxy with no evidence of an increase in \wha in its nucleus (top), and of another galaxy with an increase in \wha in its nucleus (bottom), suggesting that the latter hosts a weak AGN. The fraction of galaxies with signs of a weak AGN increases as $\wha$ approaches the 3\,\AA\ threshold.

\section{Estimating the bolometric correction for AGN from their optical spectra}
\label{bolcor}

\begin{figure*} 
  \centering
  \includegraphics[width=0.72\linewidth, trim=0 0 0 0]{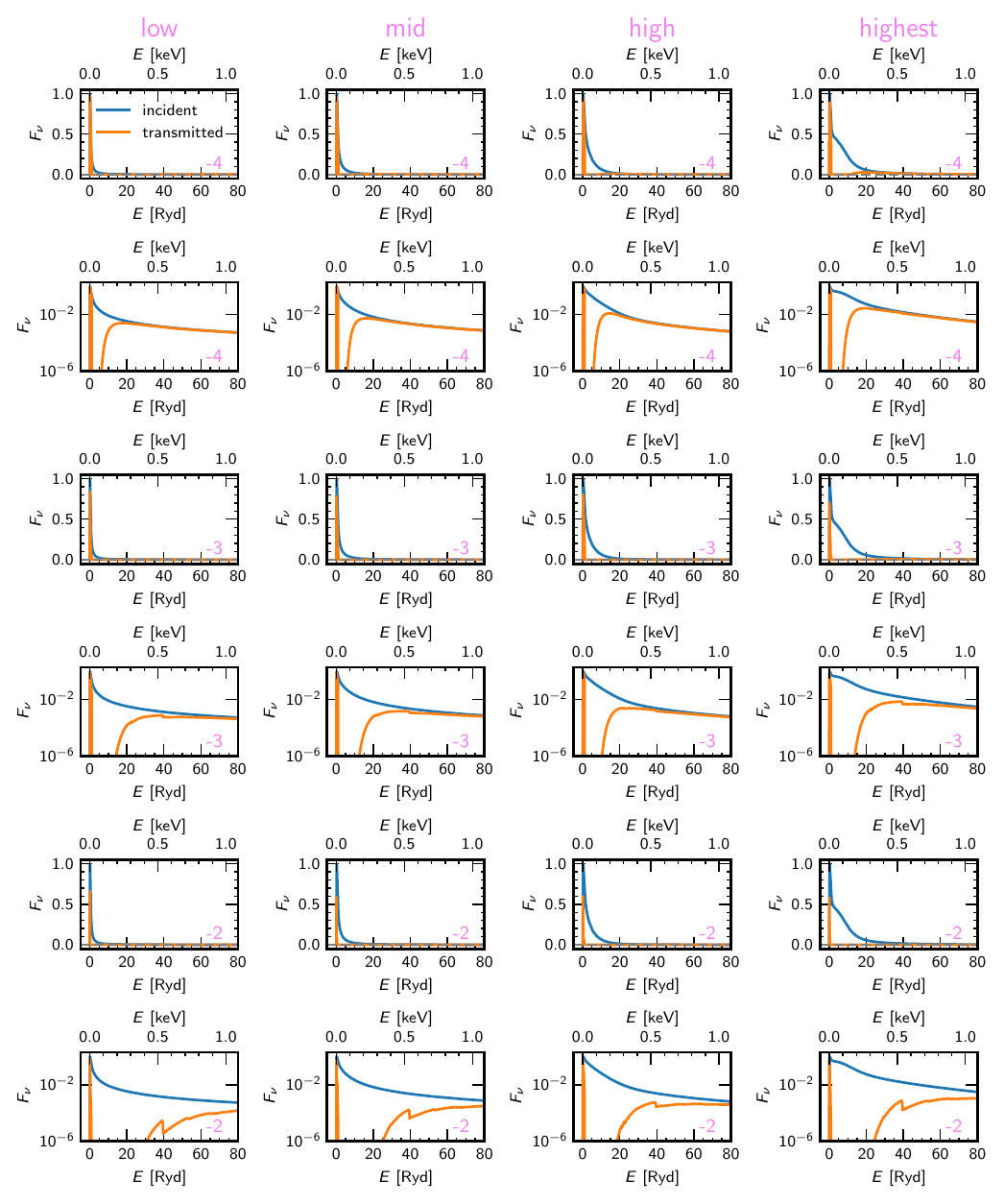} 
  \caption{Incident (blue) and transmitted (orange) SEDs for the dusty models. From left to right, we show the low, mid, high and highest SEDs (see text for details). Each SED is shown twice, with the ordinate both in linear (top) and logarithmic (bottom) scales. Numbers at the bottom right are the values of $\log U$ for each model. Models for $\log U = -2.5$ and $-3.5$ are not shown.}
  \label{fig:seds}
\end{figure*}

\subsection{AGN photoionization models}

We have run AGN models using Cloudy version 17.02 \citep{Ferland.etal.2017a}.  We constructed a grid of models using all the SEDs (low, mid, high and highest) from \citet{Ferland.etal.2020a} based on \citet{Jin.Ward.Done.2012a} and \citet{Jin.etal.2018b}. Models are run for different values of oxygen abundance ($\mathrm{O/H}$) and ionization parameter, which is defined by
\begin{equation}
  U \equiv \frac{Q_\mathrm{_H}}{4 \pi r^2 n_\mathrm{H} c},
\end{equation}
where $Q_\mathrm{H}$ is the rate of ionizing photons emitted by the source per second, $r$ is the distance from the source to the nebula, $n_\mathrm{H}$ is the volumetric density of hydrogen atoms in the nebula, and $c$ is the speed of light. In the following, we detail our model parameters.

\begin{itemize}

\item We adopt the following values for the ionization parameter: $\log U = -2.0, -2.5, -3.0, -3.5$ and $-4.0$.  We choose three values for the oxygen abundance, $12 + \log \mathrm{O/H} = 8.60$, $8.80$ and $9.02$, which roughly correspond to a sub-solar, solar and super-solar value after considering oxygen depletion onto dust grains.

\item We consider all elements that have solar abundances greater than $\log \mathrm{X/H} = -8$ and we adopt the \cite{Asplund.etal.2009a} solar photosphere abundances. All element abundances are varied in lockstep with oxygen, except for carbon and helium, for which we adopt the prescriptions from \citet{Dopita.etal.2013a}, and nitrogen, for which we use the `NHlow' relation between O/H and N/O from \citet{Zhu.Kewley.Sutherland.2023a}.

\item We assume an open plane-parallel geometry, a filling factor of one, and a constant density of $10^{3} \mathrm{\,cm^{-3}}$ (as in \citealp{Netzer.2009a}; see also \citealp{Dors.etal.2020a, Dors.etal.2023a, Binette.etal.2024a}).
\item We have run all sets of models with and without dust. Dust is added following the same procedure as in \citet{Stasinska.etal.2015a}, which is based on the works of \citet{RemyRuyer.etal.2014a} and \citet{Draine.2011b}. Note that \citet{Stasinska.etal.2015a}, \citet{ValeAsari.etal.2016a} and this work all adopt the broken power-law XCO,z case of \citealp{RemyRuyer.etal.2014a}, not the MW case as had been stated in the former paper.

\item For models with dust, we also adopt depletion factors from \citet{Dopita.etal.2013a}.


\end{itemize}

Figure~\ref{fig:seds} shows the incident and transmitted SEDs for solar-abundance models for both the dusty and dust-free case. All panels are in linear and logarithmic scale. It is important to realize that  not all photons with energies larger than 13.6 eV are absorbed. As a matter of fact none of the photons with energies above 1 KeV are absorbed. This means that the reliability of our estimation of $\Lbolmod$ relies very strongly on the choice of the input SED. This is why we chose observed SEDs, and not SEDs derived from theoretical models.

\begin{figure} 
  \centering
  \includegraphics[width=0.7\columnwidth, trim=20 10 20 10]{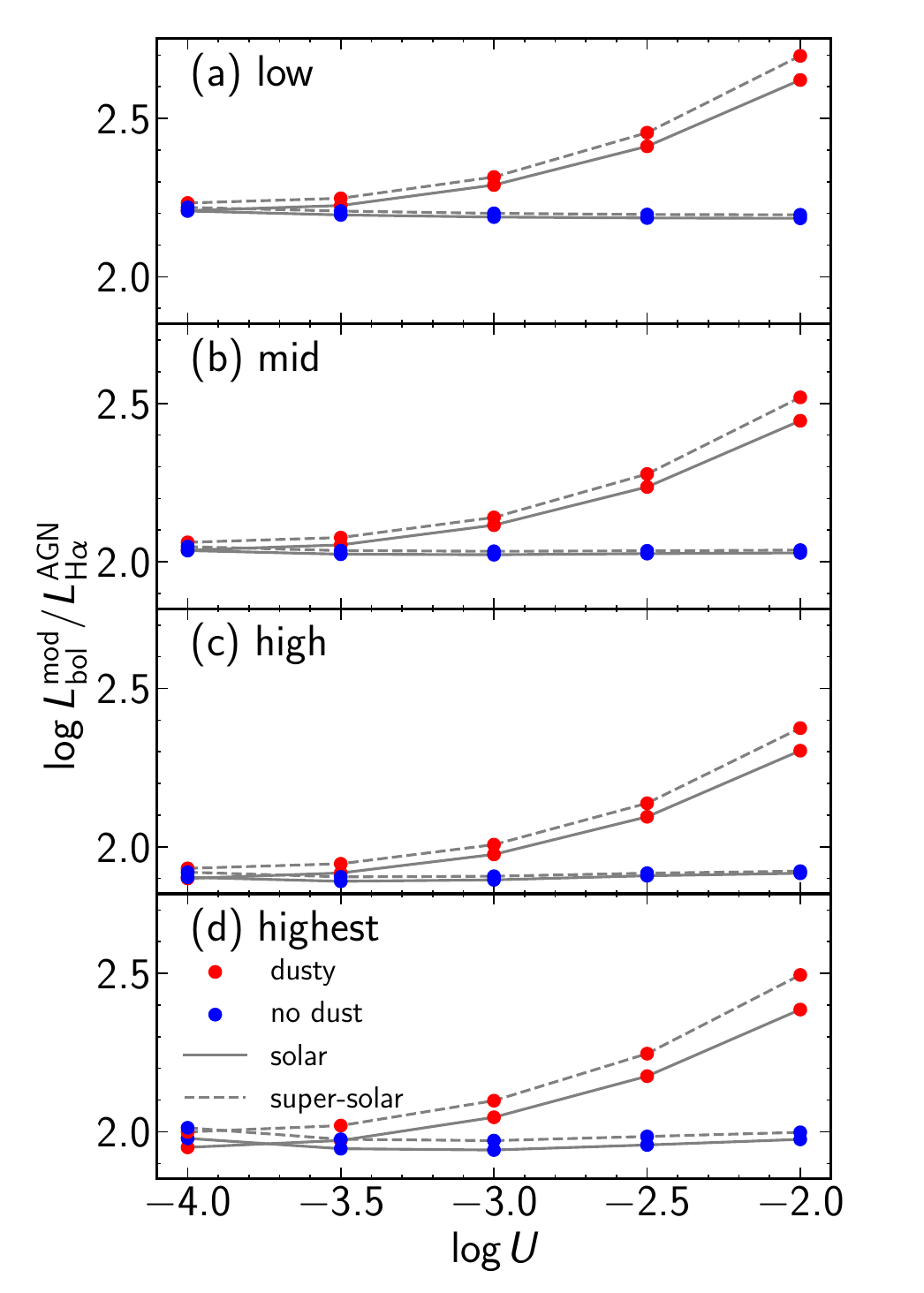}
  \caption{Model bolometric luminosity to \Ha luminosity ratio as a function of the ionization parameter $U$. The four panels show the results for all the SEDs considered here (low, mid, high and highest). Dusty models are plotted in red, and dust-free models in blue. Solid lines connect solar metallicity models, and dashed lines super-solar metallicity models.}
  \label{fig:Lbolh}
\end{figure}

Figure~\ref{fig:Lbolh} shows the bolometric-to-\Ha luminosity ratio, $\Lbolmod/\LHa$, for all SEDs for models with and without dust. \Lbolmod is calculated directly from the input SEDs; \LHa is proportional to the absorption of Lyman-photons and comes from Cloudy models. Note that, for the dust-free models, the values of $\Lbolmod/\LHa$ do not depend on $U$, and range from $\log \Lbolmod/\LHa = 1.8$ to $2.2$, depending of the SED. On the other hand, in dusty models, the values of $\Lbolmod/\LHa$ depend strongly on $U$, being higher for larger values of $U$ because in that case dust grains dominate neutral atoms for the absorption of the ionizing radiation.

\subsection{\hii region photoionization models}

Our aim is to compute \Lbol for AGNs using their dust-corrected \LHa. There are other sources in a galaxy that can ionize the gas and emit \Ha. In this work, we consider that objects falling on the `AGN wing' on the \nii/\Ha versus \oiii/\Hb diagram (the BPT diagram) have a mixture of AGN and \hii region emission. Thus we have created \hii region models considering a Starburst99\footnote{\url{https://www.stsci.edu/science/starburst99/docs/default.htm}} \citep{Leitherer.etal.1999a} SED for a 4 Myr continuous star formation with a \citet{Salpeter.1955a} initial mass function with masses ranging from $0.1$ to $120\,\mathrm{M}_\odot$, `Geneva 1994' tracks with high mass-loss rates and Pauldrach-Hillier atmospheres.
Our grid of \hii region models covers the same range in O/H and $U$ as the AGN model grid. 
We have chosen a solar-metallicity stellar population.
Nebular geometry, element abundances and depletion onto dust are treated in exactly the same manner as in the dusty AGN models. The nebular density is taken to be $10^{2} \mathrm{\,cm^{-3}}$. The criterion to stop the model calculations is when the electronic temperature reaches 200~K.

%

\begin{figure*} 
  \centering
  \includegraphics[width=0.8\textwidth, trim=0 0 0 20, clip]{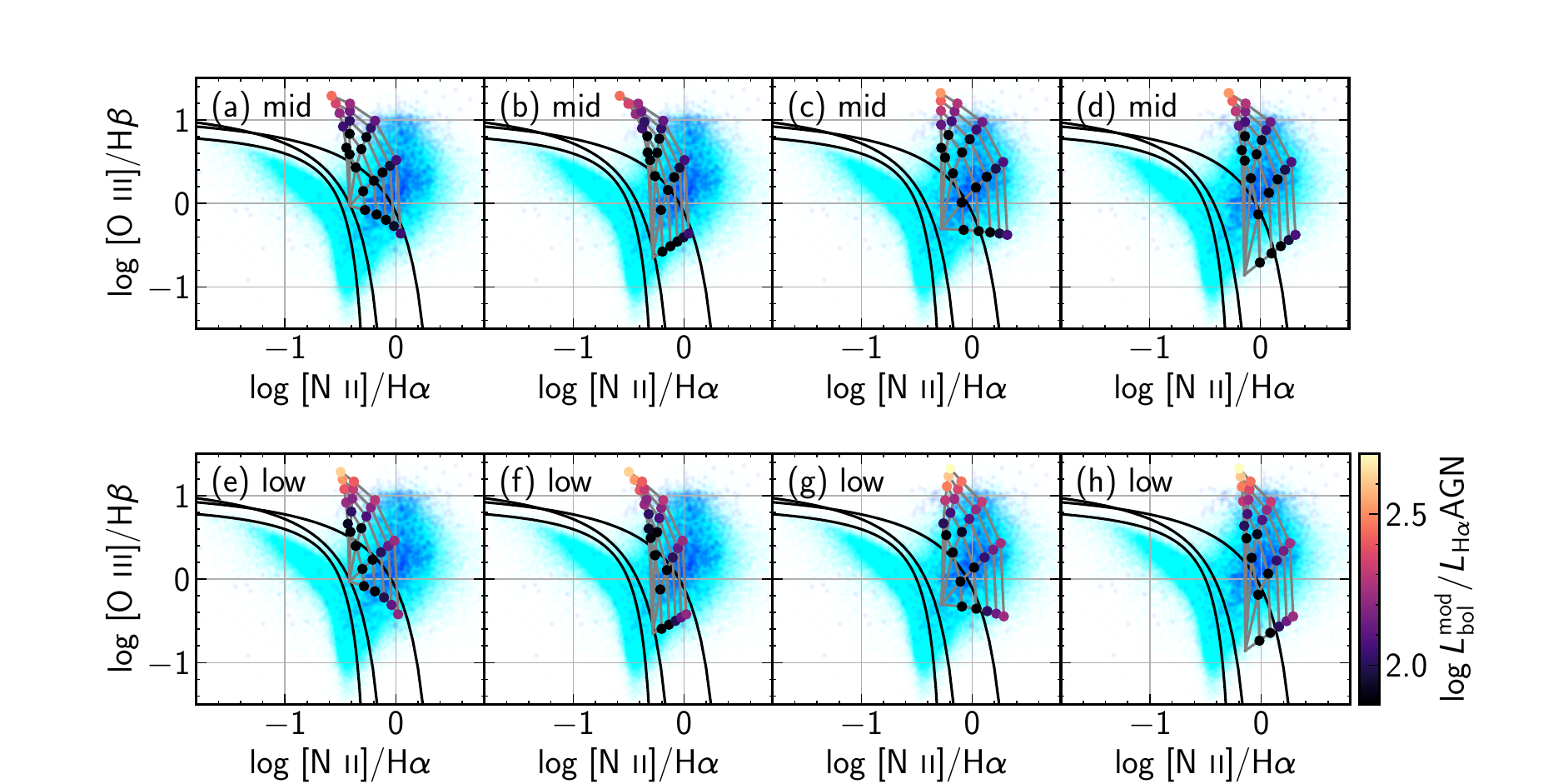} 
  \caption{BPT showing the mixing lines (solid grey lines) between AGN and \hii region models. Each panel is for a single \hii region mixed with AGN models of a given SED (indicated on the top left) spanning all values of $U$ from our grid of models.
    Points on the mixing lines mark $\eta = 20, 40, 60, 80$ and $100\%$ and are colour-coded by the values of \Lbolmod/\LHa.
    Background points in cyan are galaxies from the MGSz sample and in blue are OPARGS galaxies; curves in black are the \citet{Stasinska.etal.2006a}, \citet{Kauffmann.etal.2003c} and \citet{Kewley.etal.2001a} lines.
  }
  \label{fig:BPTmix}
\end{figure*}

\subsection{Mixing AGN and \hii region models}

\begin{figure} 
  \centering
  \includegraphics[width=0.52\linewidth, trim=120 50 50 80, clip]{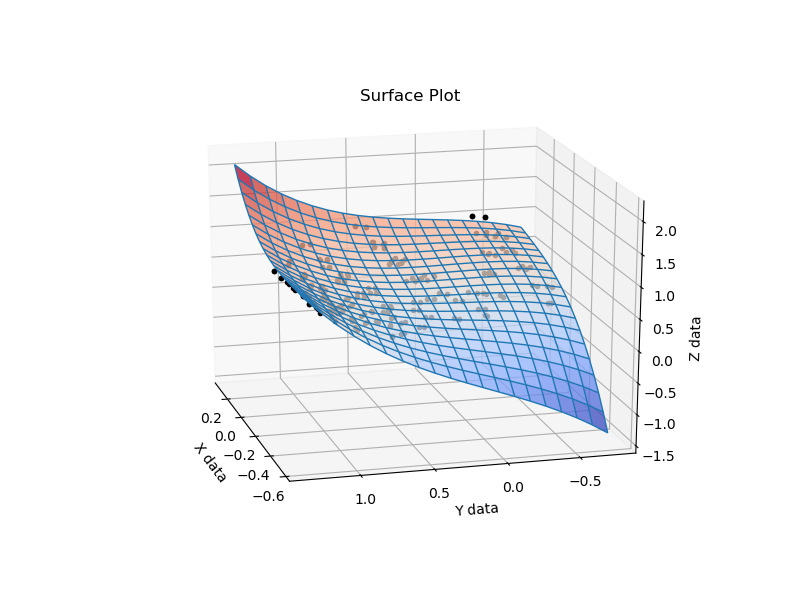}
  \includegraphics[width=0.47\linewidth, trim=0 0 50 0, clip]{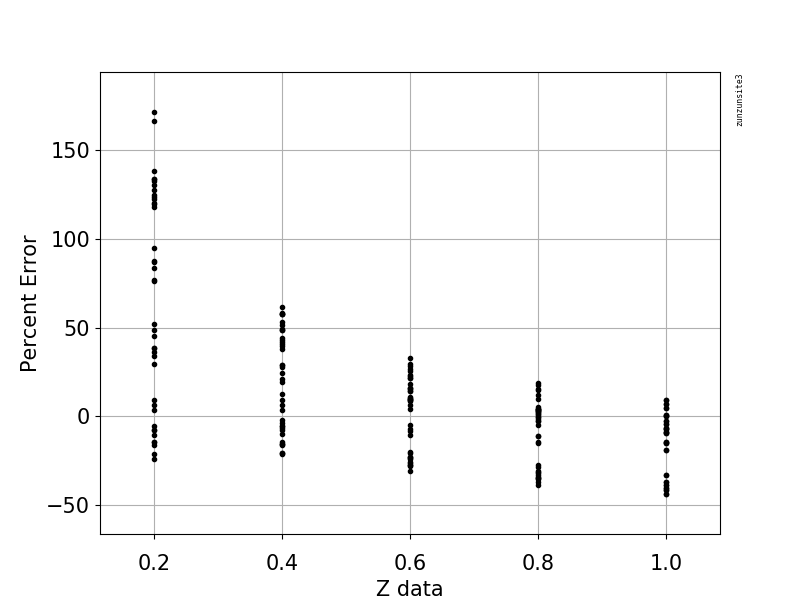}
  \caption{Results from Zunzun fitting of $\eta$. Left: surface plot for the ZunZun fitting of $\eta$; X data represents $\log\, \nii/\Ha$, Y data represent $\log\, \oiii/\Hb$, Z data represents $\eta$. Right: percentage error in the fitting as a function of $\eta$.}
  \label{fig:zunzuneta}
\end{figure}

\begin{figure} 
  \centering
  \includegraphics[width=0.52\linewidth, trim=120 50 50 80, clip]{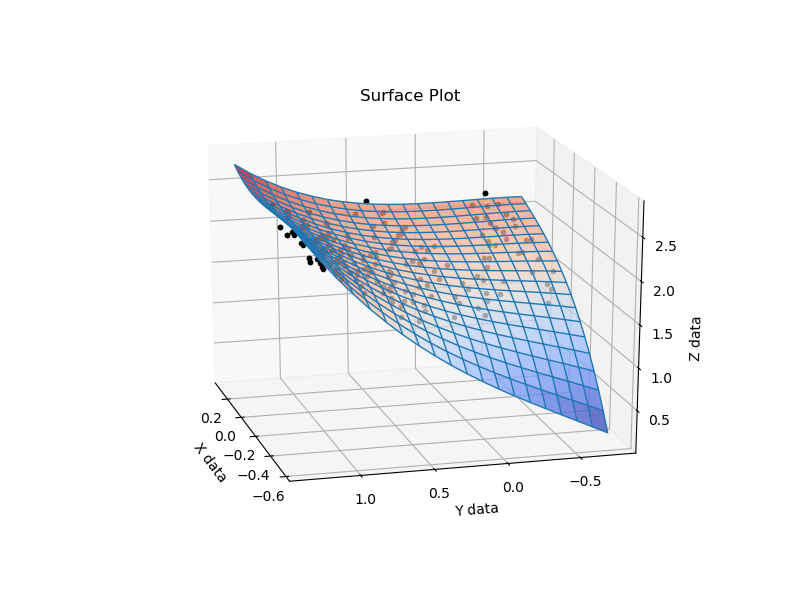} 
  \includegraphics[width=0.47\linewidth, trim=0 0 50 0, clip]{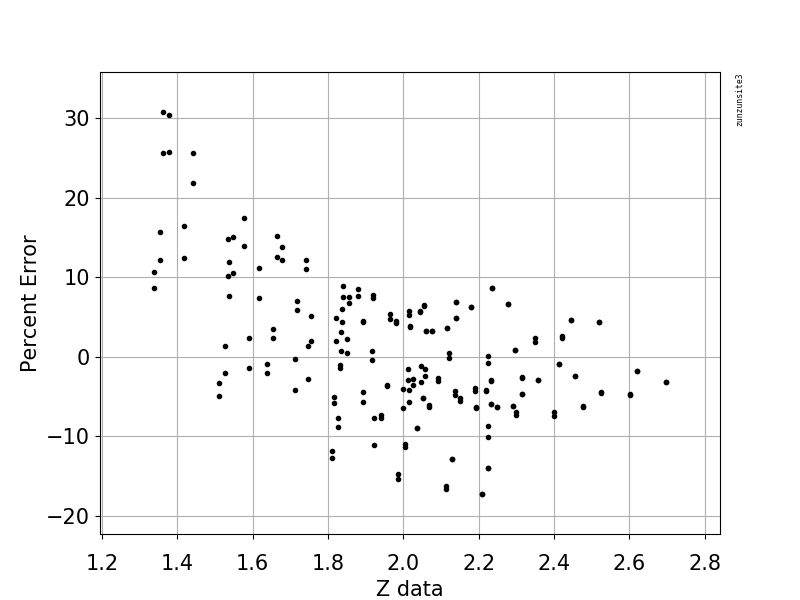} 
  \caption{Same as Fig.~\ref{fig:zunzuneta} but for $\mathrm{Z} = \log \Lbolmod/\LHa$.}
  \label{fig:zunzunlbol}
\end{figure}

\begin{figure} 
  \centering
  \includegraphics[width=.6\columnwidth, trim=20 220 350 0]{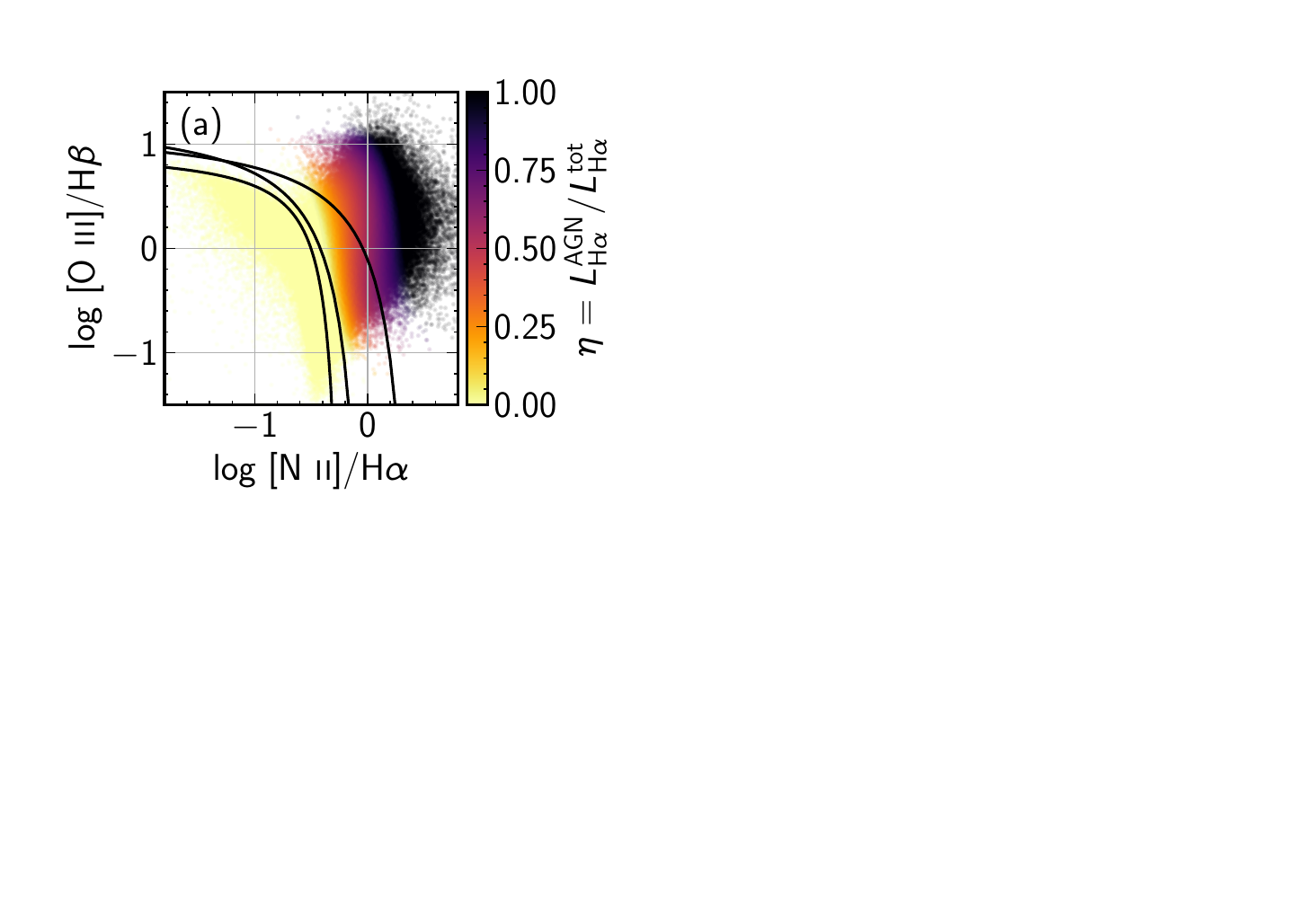} 
  \caption{Values of $\eta$ calculated for our MGSz sample using the results for the surface interpolation of all model mixing lines. Curves in black are the \citet{Stasinska.etal.2006a}, \citet{Kauffmann.etal.2003c} and \citet{Kewley.etal.2001a} lines.}
  \label{fig:BPTmixeta}
\end{figure}

\begin{figure} 
  \centering
  \includegraphics[width=\columnwidth, trim=30 240 30 0]{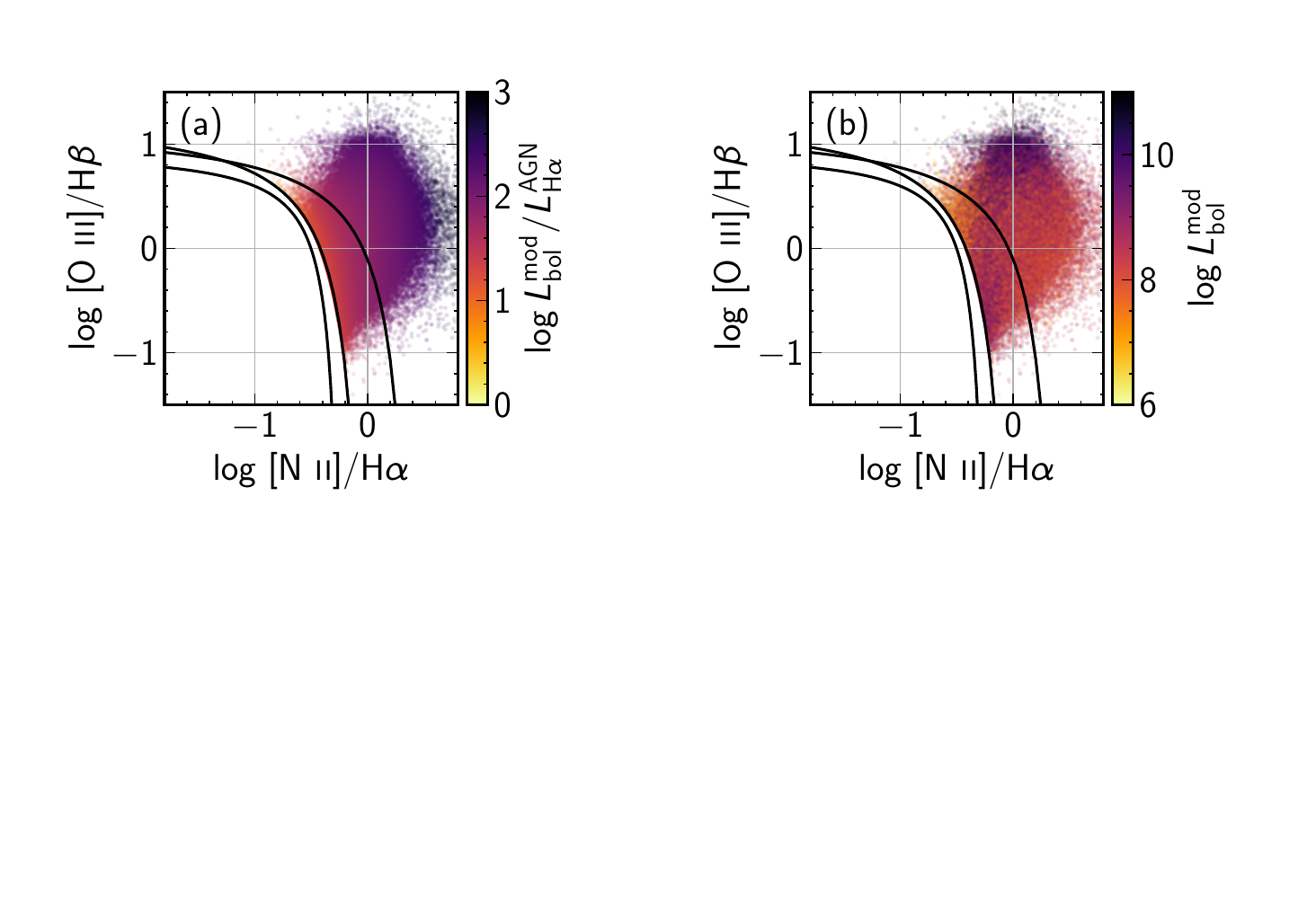} 
  \caption{Values of (a) \Lbolmod/\LHa and (b) \Lbolmod calculated for our MGSz sample using the results for the surface interpolation of all model mixing lines.}
  \label{fig:BPTmixbol}
\end{figure}

We mix AGN and \hii region models varying the proportion of \Ha luminosity coming from the AGN, parametrized by
\begin{equation}
  \eta \equiv \frac{L(\Ha)^\mathrm{AGN}}{L(\Ha)^\mathrm{AGN} + L(\Ha)^{\mathrm{H~}\small{\textsc{ii}}}}
            = \frac{L(\Ha)^\mathrm{AGN}}{L(\Ha)^\mathrm{tot}}.
\end{equation}

For the AGN models, we consider here only the dusty ones. The presence of dust in the narrow-line regions of AGNs is a highly discussed matter \citep[\eg][]{Laor.Draine.1993a, Netzer.Laor.1993a, Nagao.Maiolino.Marconi.2006a, Groves.Dopita.Sutherland.2004a, Huang.Lin.Shields.2023a, Zhu.Kewley.Sutherland.2023a}. While there have been arguments against the presence of dust inside the narrow-line region of AGNs, the strongest ones seem in favour of its presence.

Figure~\ref{fig:BPTmix} shows the loci of our mixing lines in the BPT diagram, overplotted on points representing the galaxies from the MGSz sample (in cyan) and those representing our OPARG sample (in blue).
We have selected only models for the low and mid SEDs and with solar and super-solar abundances. We mix \hii and AGN models that have the same values of O/H. Since there is no reason for the AGN and \hii region to have the same ionization parameter, we mix AGNs with all values of $U$ with \hii regions which fall near the bottom of the \hii region wing, which in practice meant selecting $\log U = -3.5$ and $-3.0$ for the \hii region models. In each panel of Figure~\ref{fig:BPTmix}, we show mixing lines radiating from a single \hii region model of a given O/H and $U$, connecting it to AGN models of the same O/H and the full range of $U$s. Points along the mixing lines mark $\eta = 20, 40, 60, 80$ and $100\%$ and are colour-coded by the values of \Lbolmod/\LHa.

Each one of the eight panels in Fig.~\ref{fig:BPTmix} defines a surface for $\eta$ and for \Lbolmod/\LHa on the BPT plane. We have then fitted a two-dimensional polynomial to represent all those surfaces at the same time.
The top panels of Figs.~\ref{fig:zunzuneta} and \ref{fig:zunzunlbol} show all the points considered and the resulting fitted surface for $\eta$ and \Lbolmod/\LHa, respectively. The bottom panels show the percentage errors of the fitting as a function of the fitted value.
The fitting was done using the ZunZun website, developed by James R.\ Phillips, and now maintained at \url{http://findcurves.com/}. 
We see that the errors in the fitting for $\log \Lbolmod/\LHa$ are less than 10\% in the majority of cases, but can be large for low values of  $\eta$.

The fitted coefficients for $\eta$ as a function of $x = \log\, \nii/\Ha$ and $y = \log\, \oiii/\Hb$ are
\begin{align}
\begin{split}  
  \eta &=0.5616 + 1.1472 x -0.0382 y -0.0850 x^2 + 0.0078 y^2
\\     &+ 2.3762 x^3 + 0.3105 y^3 -0.0539 x y + 2.4653 x^2 y
\\     &+ 0.1437 x y^2.
\end{split}
\end{align}
Figure~\ref{fig:BPTmixeta} shows how $\eta$ varies on the BPT plane. It is worth noting that on the \citet{Kewley.etal.2001a} demarcation line there are models with $\eta$ varying from $15$ to $75\%$, as had already been remarked by \citet{Stasinska.etal.2006a}.
The \citet{Kewley.etal.2001a} line was originally meant to be a `maximum starburst line'; we now know that the evolutionary tracks and stellar atmospheres then available  yielded an ionizing field that was too hard (see \citealp{Vazquez.Leitherer.2005a}).  Calling objects above the \citet{Kewley.etal.2001a} `pure AGNs', as is still routinely done in the literature,  is a misnomer given that the AGN contribution to the total \Ha emission may be as low as 20\%, in agreement with the findings of \citet{Thomas.etal.2018b}.

The fitted coefficients for $\log \Lbolmod/\LHa$ as a function of $x = \log\, \nii/\Ha$ and $y = \log\, \oiii/\Hb$ are
\begin{align}
\begin{split}  
\log \Lbolmod/\LHa &= 1.8279 + 0.9598 x + 0.1025 y -0.5880 x^2 
\\            &+ 0.1639 y^2 + 1.5134 x^3 + 0.1410 y^3 -0.6535 x y 
\\            &+ 1.2002 x^2 y -0.0656 x y^2.
\end{split}
\end{align}
Panel (a) of Fig.~\ref{fig:BPTmixbol} shows the resulting \Lbolmod/\LHa for our sample on the BPT plane. We only compute \Lbolmod/\LHa for objects above the \citet{Kauffmann.etal.2003c} line, since objects below it are dominated by star formation and the correction is highly uncertain. Panel (b) shows \Lbolmod for the same objects; note how objects at the top of the wing are special, just like highlighted in Sect.~\ref{BPT}. We remind the models have been run for a covering fraction of 1, but, as explained in Sect.~\ref{boloparg}, we recommend taking the covering fraction $0.65$, so that $\Lbol = \Lbolmod/0.65$.

\subsection{The impact of HOLMES}

In determining the bolometric luminosity, we have supposed that the energy emitted in the emission lines comes only from \hii regions and the AGN and that nothing comes from ionization by HOLMES. In objects having  values of \wha just slightly above 3\AA, this may not be entirely true. However it is difficult to estimate the proportion of HOLMES without a dedicated stellar population analysis for each object. But the effect should not be much different from what we computed above, since, if HOLMES contribute to the emission lines, the values of \oiii/\Hb and \nii/\Ha to be used are not the observed ones but values discounting the effect of HOLMES, which would lead to higher values of the bolometric correction to apply to the \Ha luminosity. However, the \Ha luminosity to which the correction should be applied is  the observed \Ha luminosity after correction for extinction and for the contribution of the HOLMES. The final result is that HOLMES do not strongly affect the determination of the bolometric luminosity. 

\end{appendix}

\label{lastpage}

\end{document}